\documentclass[aps,prb,floatfix,reprint]{revtex4-1}
\usepackage{graphicx}
\usepackage{subfigure}
\usepackage{amsfonts}
\usepackage{amsmath}
\usepackage{amssymb}
\usepackage{subfigure}
\usepackage{soul}

\usepackage{xcolor,soul}

\begin{document}

 \title{Nonequilibrium current-induced forces caused by quantum localization: 
 Anderson Adiabatic Quantum  Motors.}
 
\author{Lucas J. Fern\'andez-Alc\'azar$^{1}$}
\author{Horacio M. Pastawski$^{1}$}
\author{Ra\'ul A. Bustos-Mar\'un$^{1,2}$}
\thanks{Corresponding author: \texttt{rbustos@famaf.unc.edu.ar}}

\affiliation{$^1$Instituto de F\'{\i}sica Enrique Gaviola and Facultad de
Matem\'{a}tica Astronom\'{\i}a, F\'{\i}sica y Computación, Universidad Nacional de
C\'{o}rdoba, Ciudad Universitaria, C\'{o}rdoba, 5000, Argentina}

\affiliation{$^2$Facultad de Ciencias Qu\'{\i}micas, Universidad Nacional de C\'{o}rdoba, Ciudad
Universitaria, C\'{o}rdoba, 5000, Argentina}

\begin{abstract}
In recent years there has been an increasing interest in nanomachines.
Among them, current-driven ones deserve special attention as quantum effects can play a significant role there. Examples of the latter are the so-called adiabatic quantum motors.
In this work, we propose using Anderson's localization to induce nonequilibrium forces in adiabatic quantum motors. We study the nonequilibrium current-induced forces and the maximum efficiency of these nanomotors in terms of their respective probability distribution functions.
Expressions for these distribution functions are obtained in two characteristic regimes: the steady-state and the short-time regimes. Even though both regimes have distinctive expressions for their efficiencies, we find that, under certain conditions, the probability distribution functions of their maximum efficiency are approximately the same. Finally, we provide a simple relation to estimate the minimal disorder strength that should ensure efficient nanomotors.
\end{abstract}
\maketitle

\section{Introduction}

In the last decades, control and fabrication of nanoelectromechanical systems have had a huge boost 
enabled by
the advances in our control over matter at the nanoscale and 
stimulated by
the applications they promise us.\cite{Craighead2000_NEMS,roukes2001_NEMS}
For example, they could be used for harvesting different energy sources at the nanoscale, 
cooling nanodevices, or even for building complex nanomachines. \cite{Craighead2000_NEMS,roukes2001_NEMS,MolcecMotorsGral,Dundas,Bailey08,
bustos2018,McEniry09current_cooling,galperin2009cooling,
arrachea2012refrigerator,FPB17,FastNM,nanocar,
NatureNano11,chiaravalloti2007rack} 
Moreover, among all the proposed mechanisms that can be used to control nanomachines, the use of electric currents is particularly appealing due to its compatibility with current technologies involved in modern electronics circuits.

There are several interesting theoretical and experimental examples in the literature of current-driven nanomachines. \cite{Dundas,Bailey08,nanocar,MolcecMotorsGral,NatureNano11,chiaravalloti2007rack,FastNM,nanocar}
A remarkable proposal, which could take advantage of quantum effects at the nanoscale,
is the so-called adiabatic quantum motor.
\cite{AQM13,FBP15,arrachea2015,Silvestrov2016,Liliana_Onsager,FPB17,calvo2017}
This consists of a mechanical device, typically nanometric,
capable of being moved by a ``wind''\cite{landauer74driving} of quantum particles.
In the adiabatic quantum motors, the quantum nature of the driving particles 
can be exploited to boost the performance of such motors.
This is the case of, e.g., adiabatic quantum motors based on the Thouless pump.\cite{AQM13,arrachea2015,Thouless83,Zhang,FPB17} 
There, a mobile piece induces a periodic potential on a conductor where the movement of the piece translates into a displacement of the potential.
The periodic potential induces a gap in the dispersion relation of the electrons.
Carriers with energy within this gap cannot cross the conductor and, thus, they suffer a backscattering process with the consequent transfer of momentum to the mechanical piece.
Then, as only backscattered electrons contribute to the transfer of momentum, 
low transmittances are the key to the efficiency of the Thouless adiabatic quantum motor.
However, other quantum effects can also reduce transmittances and increase the efficiency of adiabatic quantum motors.

For long conductive wires, impurities or defects are commonly present in experimental samples, 
inducing disorder in the potential energy perceived by the propagating electrons. 
For coherence lengths large enough, Anderson's localization of the electrons' wave functions arises.
While localized states are generally taken as a drawback for quantum transport,
\footnote{
It has been shown that localization can improve charge pumping.
See for example Refs. \cite{chern2007} and \cite{ingaramo2013}
}
for adiabatic quantum motors they can turn into a welcomed feature,
if they are caused by impurities in the movable part of the device as we will see.
In that case, the exponentially reduced transmittances induced by localization can translate into an increased efficiency of the nanomotors.
\footnote{
While this work was under the reviewing process, we found a work with a related idea, but about the possibility of using many-body-localization to make an engine.
N. Y. Halpern, C. D. White, S. Gopalakrishnan, and G. Refael . MBL-mobile: Many-body-localized engine. arXiv:1707.07008 (2018)
}

In this work, we assess the possibility of using Anderson's localization to induce nonequilibrium forces in adiabatic quantum motors. We study this kind of devices, which we call ``Anderson adiabatic quantum motor'' (AAQM), in terms of probability distribution functions of their properties, discussing the conditions that would warrant their proper functioning.

The work is organized as follows. 
In Sec. \ref{sec:CIFs} we derive the general equations of nonequilibrium current-induced forces (CIFs) for the case of nanodevices where there is a shift of the potential energy sensed by the electrons.
We derive the expressions of the CIFs by using a scattering-matrix approach but also from 
intuitive arguments based on momentum conservation.
In Sec. \ref{sec:CIFs_Anderson} we evaluate the CIFs for the particular case of quasi-unidimensional disordered potentials by means of the Anderson's model of disorder. We also compare the theoretical results of CIFs with numerical simulations.
In Sec. \ref{sec:efficiency} we discuss the efficiency of AAQMs distinguishing two dynamical regimes of interest:
the short-time and the steady-state regimes (sections \ref{sec:efficiency_transient} and \ref{sec:efficiency_Steady} respectively).
We derive expressions for the probability distribution function of the optimal efficiency of the nanomotors. We also identify the regime that should ensure that most of the AAQMs will be efficient.
Finally, in Sec. \ref{sec:Conclusions} we summarize our main results and discuss the possible extensions and consequences of them.

\section{Current-induced forces\label{sec:CIFs}}
\begin{figure}[ptb]
\begin{center}
\includegraphics[width=0.45\textwidth]{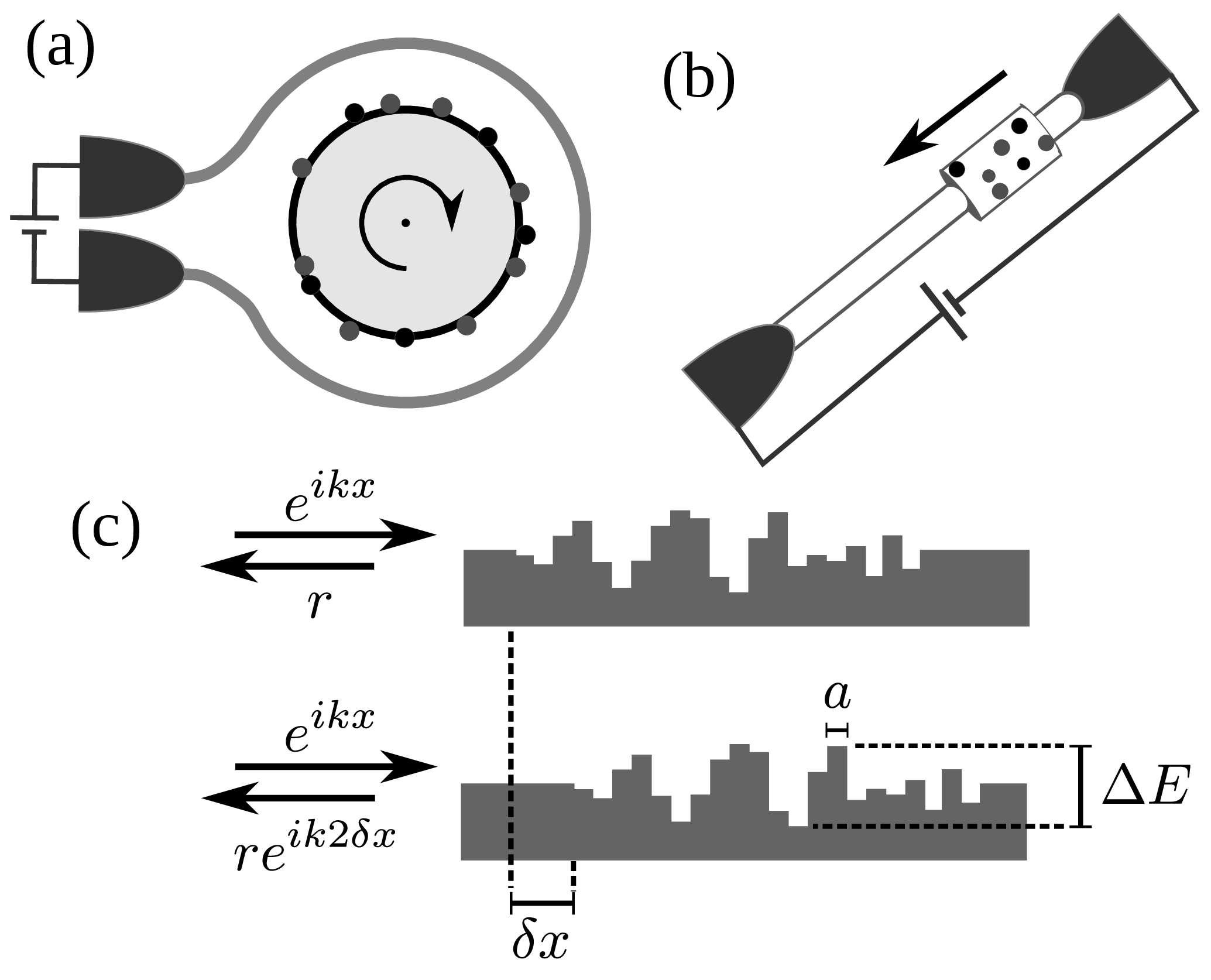}
\end{center}
\caption{
(a), (b) Examples of AAQMs. Panel (a) stands for a conductive wire coiled around a rotating piece with randomly placed charges.
Panel (b) represents a multiwall nanotube where the inner nanotube is longer and the outer one has random impurities. The outer nanotube is supposed to be free to move along the guide given by the inner nanotube.
Panel (c) schematizes the changes of the potential energy (characterized by $a$ and $\Delta E$) sensed by the electrons inside the conductive wire of example (a) or the inner nanotube of example (b). There, a displacement of the potential by $\delta x$ produces a phase change of $2k\delta x$ on the reflection coefficient.
}
\label{fig:potential}
\end{figure}
In the present work, we deal with devices like those depicted in Fig. \ref{fig:potential}.
There, the flow of electrons, induced by a voltage bias between two reservoirs, moves a rotating piece [the rotor in Fig. \ref{fig:potential} (a)] or push an artifact along a track [the shuttle in Fig. \ref{fig:potential} (b)]. The physical reason why those nanoelectromechanical devices work can be readily understood in terms of linear momentum conservation.
Indeed the mathematical expressions for the current-induced forces (CIFs) can be derived intuitively solely based on that.

For simplicity, we will not consider the electron spin and we will neglect all equilibrium
CIFs, e.g., those forces due to scattering at the system-lead boundaries.\cite{FPB17}
Let us take the case of the shuttle depicted in panel $(b)$ of Fig. \ref{fig:potential}.
For the rotor, panel $(a)$ of Fig. \ref{fig:potential}, the following arguments will remain since the potential sensed by the electrons changes with $\theta$ (the angle that sets the position of the rotor) in almost the same way as how the potential in the shuttle changes with $x$ (the coordinate that sets the position of the shuttle). See appendix \ref{App:potential}.
In panel $(b)$ of Fig. \ref{fig:potential}, the presence of the shuttle perturbs the potential sensed by the electrons moving along the wire yielding a Hamiltonian $\hat{H}_e(x)$.
The dependence of $\hat{H}_e$ with the relative position of the moving piece $x$ 
results from the displacement of the potential.
The net force produced by the current comes from the interaction of the electrons with this potential.
Electrons injected from a reservoir $\alpha$ can be reflected with probability $R_{\alpha}(\varepsilon)$.
Due to linear momentum conservation, the electrons reflected transfer a momentum $\Delta p_{\alpha}$ to the shuttle.
The number of electrons per unit time coming from a channel $\alpha$ with energy between $\varepsilon$ and $d\varepsilon$ is 
\begin{equation}
(f_{\alpha} N_{\alpha} d\varepsilon/2) v_{\alpha},
\end{equation}
where the first quantity in parenthesis is the number of electrons per unit length moving towards the system, and $v_{\alpha}(\varepsilon)$ is the velocity of the carriers. The function $f_{\alpha}(\varepsilon)$
is the occupation probability, $N_{\alpha}(\varepsilon)$ is the density of states per unit length, and the factor $(1/2)$ comes from counting only the electrons traveling in one of the two possible directions. 
Now, the net force sums contributions of electrons coming from all reservoirs and possible energies (integrating over $\varepsilon$). The result is
\begin{equation}
F_{x}=\sum_{\alpha}\int f_{\alpha}(\varepsilon)\frac{N_{\alpha}(\varepsilon
)}{2}v_{\alpha}(\varepsilon)R_{\alpha}(\varepsilon)\Delta p_{\alpha
}(\varepsilon)\mathrm{d}\varepsilon.
\label{eq:fuerza_momento}
\end{equation}
Noticing that the density of states and the group velocity compensates precisely $N_{\alpha}\equiv2/(hv_{\alpha})$, yields the simpler form
\begin{equation}
F_{x}=\sum_{\alpha}\int\frac{1}{h}f_{\alpha}(\varepsilon)R_{\alpha
}(\varepsilon)\Delta p_{\alpha}(\varepsilon)\mathrm{d}\varepsilon,
\label{eq:force}
\end{equation}
where $h$ is the Planck's constant.
Similar expressions have been derived using heuristic or semiclassical arguments.\cite{Peierls1975force,Saenz01,Bailey08}
In the following, we will arrive at the same expression through formal quantum arguments.

If the moving piece of a nanodevice is large enough it is usually a good approximation 
to treat the system under 
the nonequilibrium Born-Oppenheimer approximation \cite{bennett2010,thomas2012}, or Ehrenfest approximation. \cite{diventra2000,horsfield2004,todorov2010}
In these, the dynamics of the electronic and mechanical degrees of freedom are well separated in time and the mechanical degrees of freedom can be treated classically.
The dynamics of the mechanical nanodevice is governed by the mean value of the CIFs exerted by the quantum particles over the classical degrees of freedom $\vec{x}$.
The expectation value of the force operator $F_x$ is given by
\begin{equation}
F_x = -\left < \frac{\partial \hat{H}_e}{\partial x}\right > = \mathrm{tr} \left[ \mathrm{i} \hbar \frac{\partial H_e}{\partial x} {\cal G}^{<} (t,t) \right], 
\end{equation} 
where $\mathrm{tr} \left[ \bullet \right]$ is the trace, $\hat{H}_e$ is the electronic Hamiltonian, and
${\cal G}^{<} (t,t)$ is the lesser Green's function in the Keldysh-Kadanoff-Baym formalism\cite{haug2008,GLBE2}, which evaluated at equal times is proportional to the density matrix.
The above expression can be handled to be fully written in term of the scattering matrix $S$,\cite{bennett2010,bode2011,bode2012,thomas2012} resulting in
\begin{equation}
F_{x}=\sum_{\alpha}  \int \frac{\mathrm{d}\varepsilon}{2\pi\mathrm{i}}f_{\alpha}(\varepsilon
)\left\{  S^{\dagger}\frac{\partial S}{\partial x}\right\}_{\alpha \alpha}.
\label{eq:force_scatt}
\end{equation}
For one dimensional systems, the scattering matrix of spinless noninteracting particles is a $2 \times 2$ matrix.
The most general unitary scattering matrix of this type can always be written as
\begin{equation}
S=
\begin{pmatrix}
r & t^{\prime}\\
t & r^{\prime}
\end{pmatrix}
=e^{\mathrm{i}\chi}
\begin{pmatrix}
e^{\mathrm{i}\theta}\cos\beta & \mathrm{i}e^{-\mathrm{i}\phi}\sin\beta\\
\mathrm{i}e^{\mathrm{i}\phi}\sin\beta & e^{-\mathrm{i}\theta}\cos\beta
\end{pmatrix}
, \label{eq:S_matrix}
\end{equation}
where $r$ and $t$ are respectively the reflection and transmission amplitude coefficients; $\chi\in\lbrack0,\pi)$ is a global arbitrary phase that depends on the choice of the origin for the two channels; $\theta\in\lbrack0,2\pi)$ varies when the scatterer is shifted; $\beta\in\lbrack0,\pi/2]$ 
determines the module of reflection and transmission coefficients; and $\phi\in\lbrack0,2\pi)$ only becomes relevant for quantum pumping\cite{brouwer1998} (or CIFs) under the presence of a vector potential varying with $x$. See Ref. \onlinecite{Avron2004} for a deeper discussion.

In this work, we are only considering systems without magnetic fields ($t_{LR}=t_{RL}\Rightarrow\phi=0$).
Then, the quantity within braces in Eq. \ref{eq:force_scatt} results in 
\begin{equation}
\left \{ S^{\dagger}\partial_x S\right \}_{LL/RR} = i\left(\partial_x \chi\right) \pm i\left(\partial_x \theta\right)R
\end{equation}
where we used $S^{\dagger}S=I$, and $R=\cos^{2}\beta$.
The equilibrium force $F^{eq}$ [defined for the average Fermi energy 
$f_{0}=\left(f_{L}+f_{R}\right)/2$], and the nonequilibrium force $F^{ne}$ (where $F=F^{eq}+F^{ne}$) result in\cite{FPB17} 
\begin{eqnarray}
F^{eq} & = &  \int\frac{d\varepsilon}{\pi}f_{0}\left(\partial_x \chi\right), \notag \\
F^{ne} & = &  \int\frac{d\varepsilon}{2\pi} \left(\partial_x \theta\right)R\left(f_{L}-f_{R}\right).
\end{eqnarray}
Now using $d\theta=2kdx=2kr_{R}d\varphi$, $\pm2\hbar k=\Delta p_{L/R}$, and neglecting equilibrium forces, one arrives to Eq. \ref{eq:force}.
As discussed in Ref. \onlinecite{Avron2004} section 2.3, a change in $\chi$ is related to a variation of the occupation of the system. Therefore, the interpretation of the forces acting on the analyzed systems is the following: equilibrium forces come from changes in the occupation of the system, while nonequilibrium forces come from momentum conservation of the scattered electrons.
As mentioned before, for simplicity we will neglect equilibrium forces in the treatments of both the rotor and the shuttle.
\footnote{
As in the case of the Thouless motor\cite{FPB17}, in the rotor, weak potentials and the smoothening of the system's edges make $F^{eq}$ small as compared with $F^{neq}$. For the shuttle, $F^{eq}$ comes from imperfections of the device, which we are neglecting in the present work.
}

For a small bias voltage and low temperatures, we can simplify even further Eq. \ref{eq:force}, yielding
\begin{equation}
F  =(1-T)\frac{k_F}{\pi}\delta\mu ,
\label{eq:fuerza_lineal}
\end{equation}
where $\delta\mu =\mu_{L}-\mu_{R}$, with $\mu_{L}$ and $\mu_{R}$ being the left and right chemical potentials of the reservoirs respectively, and $k_F=k(\varepsilon_F)$ is evaluated at the Fermi energy $\varepsilon_F$. 
An average value of $T$ can also be used in Eq. \ref{eq:fuerza_lineal} if transmittances varies significantly in the energy range between $\mu_L$ and $\mu_R$.

The total pumped charge $Q$ associated to a displacement $L$ of the shuttle or the rotor can be obtained by using the Onsager's relation between the pumped current and the 
nonequilibrium part of the CIFs \cite{Cohen,AQM13,Liliana_Onsager,FBP15,FPB17,calvo2017}
\begin{equation}
 \left . \frac{\partial F}{\partial \left ( \delta \mu \right )} \right |_{eq} = 
 \left . \frac{\partial I}{\partial \dot x} \right |_{eq} \label{eq:Onsager}
\end{equation}
where $I$ is the current. 
Multiplying both sides of Ec. \ref{eq:Onsager} by $\delta \mu$ and integrating over the trajectory of the system (the integration $\int d x$ is not necessarily carried out over the period of a cyclic motion) results in
\begin{equation}
W-W^{eq}=(Q/e)\delta \mu, 
\label{eq:work_Onsager}
\end{equation}
Here, $e$ is the electron's charge, $W$ is the total work done by the CIFs, and $W^{eq}=\int F^{eq} d x$. $F^{eq}$ is the equilibrium component of $F$, i.e. $F \approx F^{eq}+\partial_\mu F \delta \mu$ in the limit of small voltages. The total charge pumped by the motion of the system, $Q$, can be calculated by using Eq. \ref{eq:work_Onsager} and assuming $F^{eq}\approx 0$, giving
\begin{equation}
 Q=e(1- \left < T \right >_x ) \frac{L k_F}{\pi}.
 \label{eq:pumped_charge}
\end{equation}
where $\left < T \right >_x$ is the average value of $T$ along the trajectory, $\left < T \right >_x = \int_0^L T dx/L$.
Note, that, for the case of the shuttle, $T$ does not change during the trajectory ($\left < T \right >_x$ and $T$ are the same), but for the rotor, there can be differences. This will affect the probability distribution function of $T$ and $\left < T \right >_x$, $P(T)$ and $P(\left< T \right >_x)$ respectively. We will address this point later in section \ref{sec:efficiency_Steady} and in appendix \ref{App:PT}.
Similar expressions to Eq. \ref{eq:pumped_charge} have been previously reported in literature \cite{Cohen2005,Avron2004}.
Notice that in Eqs. \ref{eq:fuerza_lineal} and \ref{eq:pumped_charge} the particularities of the potential profile enters only through the transmittance. Thus, the expressions are quite general.

From eqs. \ref{eq:force} and \ref{eq:fuerza_lineal}, it results evident that total reflection of electrons is crucial to maximize the force. 
In the Thouless adiabatic quantum motor studied in Refs. 
\onlinecite{AQM13,arrachea2015,Thouless83,Zhang,FPB17} 
the high performance of the proposed adiabatic quantum motor is a consequence of a reflection coefficient exponentially close to 1, as result from a precise periodicity of the potential.
Disordered unidimensional systems also present almost zero transmittance, with the advantage that much less control is required for the realization of the device. However, they have the disadvantage that they are random by nature.
In the following, we will study how the stochastic feature of nanoelectromechanical devices based on Anderson's localization affects their performance.

\section{Current-induced forces in the Anderson's model.\label{sec:CIFs_Anderson}}

Let us consider the case of electrons that move along a wire of length $L$ but whose potential energy is stochastic.
The aleatory nature of this potential can be due, e.g., to the proximity of randomly placed impurities on the surface of the rotor or the shuttle. As we know from the pioneer works of P. W. Anderson,\cite{Anderson78} the disorder in unidimensional or quasi-unidimensional systems causes the localization of eigenstates.
This quantum phenomenon can be understood as a breakdown of extended states where an eigenfunction of the system, $\psi$, can be roughly described by an exponential function as
$\psi(x) \propto e^{-|x-x_c|/\xi}$ 
for $x\rightarrow \pm \infty$, where $\xi$ is the localization length and $x_c$ is a localization center.\cite{PWA83,PSW85_breakdown,KMK1993}
The transmittance $T$ of such systems connected to reservoirs should depend exponentially on the ratio between $L$ and $\xi$ as
\begin{equation}
T\sim\exp\left(  -2/\widetilde{\xi}\right), \label{eq:transmittance}
\end{equation}
where $\widetilde{\xi}$ is the reduced localization length, $\widetilde{\xi}=\xi/L$.

For systems in a strong localized regime, $1/\widetilde{\xi} \gg 1$, the factor $(1-T)$ in Eq. \ref{eq:fuerza_lineal} can be taken as 1. In that case, the nonequilibrium CIFs take their maximum value, $F=k_F\delta\mu / \pi$.
However, a complete description of the properties of these stochastic systems 
should be given in terms of probability distribution functions. 
In our case, first we will be interested in describing $P(\widetilde{\xi})$, the probability distribution function of the reduced localization length. This function depends, in principle, on the model used to describe the disorder.

The most extensively studied model of disordered one dimensional systems is the Anderson's model.
There, the wire is described by a tight-binding chain of length $L=N a$, where $a$ is the lattice constant and $N$ the number of sites. The Hamiltonian in this case is
\begin{equation}
\hat{H}_{}=\sum\limits_{n=1}^{N}\left\{ E_{n}\hat{c}_{n}^{\dagger }\hat{c}
_{n}^{{}}-V \left[ \hat{c}_{n}^{\dagger }\hat{c}_{n-1}^{{}}+\hat{c}
_{n-1}^{\dagger }\hat{c}_{n}^{{}}\right] \right\} , \label{eq:Hs}
\end{equation}
where $\hat{c}_{n}^{\dagger}$ and $\hat{c}_n$ are the creation and annihilation operators at site $n$, and $V$ is the hopping parameter.
The disorder is modeled by random site's energies, $E_n$, which are chosen from a uniform random distribution with 
$|E_n| \le \frac{\Delta E}{2}$.
At its edges, the wire is connected to leads. This adds a self energy $\Sigma(\varepsilon)$  at the local energy $E_{1(N)}$ of the effective Hamiltonian\cite{PM01,FPB17} 
\begin{equation}
\hat{H}_{\mathrm{eff}}=\hat{H}_{}+\Sigma (\varepsilon )[\hat{c}_{1}^{\dagger }\hat{c}^{\ }_{1}+\hat{c}_{N}^{\dagger }\hat{c}^{\ }_{N}], \label{eq:H_eff}
\end{equation}
 where
\begin{equation}  
\Sigma (\varepsilon ) = \lim_{\eta \rightarrow 0^+} \frac{\varepsilon+\mathrm{i}\eta }{2}
-\mathrm{sgn}(\varepsilon)\sqrt{\left( \frac{\varepsilon+\mathrm{i}\eta }{2}\right)
^{2}-V^{2}}. \label{eq:Sigma}
\end{equation}
In this model, the parameter $a$ can be interpreted as the typical length of the defects while $\Delta E$ accounts for the width of the distribution function of their energy. If required, the hopping parameter $V$ can be obtained from the discretization of the Schr\"odinger equation in the continuous, $V=\hbar^2/2 m_e a^2$ where $m_e$ is the mass of the electron.\cite{PM01}

Within the Anderson's model there are different regimes of disorder. In this work, we will focus only on the weak disorder regime, i.e., $\Delta E\ll V$. There, the probability distribution function of $\widetilde{\xi}$ is well described by\cite{KMK1993}
\begin{equation}
 P(\widetilde{\xi}|\widetilde{\xi}_0) \propto \frac{1}{\widetilde{\xi}} \cdot \exp \left( -\frac{(1/\widetilde{\xi} - 1/\widetilde{\xi}_0)^2}{2/\widetilde{\xi}_0} \right),
 \label{eq:P_gamma}
\end{equation}
where $P(\widetilde{\xi}|\widetilde{\xi}_0)$ depends parametrically on $\widetilde{\xi}_0$ and 
the energy of the electrons $\varepsilon$ should accomplish $|\varepsilon|<2V$.
\footnote{
In all the figures of this manuscript, we used Eq. \ref{eq:P_gamma} to calculate the probability distribution 
function of the properties of interest. For other types of disorder, alternative expressions for 
Eq. \ref{eq:P_gamma} may hold. However, the expressions for the probability distribution 
functions of the force (Eq. \ref{eq:P_F}) and the efficiency (Eq. \ref{eq:P_eta_AQM}) are written in such a way that they are independent of the particular form of the probability distribution function of $\xi$.
}
The dimensionless parameter $\widetilde{\xi_0}$ accounts for the disorder-relevant microscopic details of the system
\footnote{
The ensemble described by the probability distribution functions may consist of different realizations of the system but may also come from evaluating the same system at different Fermi energies, which should involve distinct sets of localized states.
}
and can be obtained from\cite{KMK1993}
\begin{equation}
\frac{1}{\widetilde{\xi_0}}=\left(\frac{L}{a}\right) \frac{(\Delta E/V)^{2}}{96\left(1- \left(\frac{\varepsilon}{2V} \right)^{2}\right)} = \frac{L~\Delta E^{2}a}{24 v_F^2 \hbar^2},
\label{eq:xi_0}
\end{equation}
where $v_F$ is the Fermi velocity.

\begin{figure}[ptb]
\begin{center}
\includegraphics[width=0.45\textwidth]{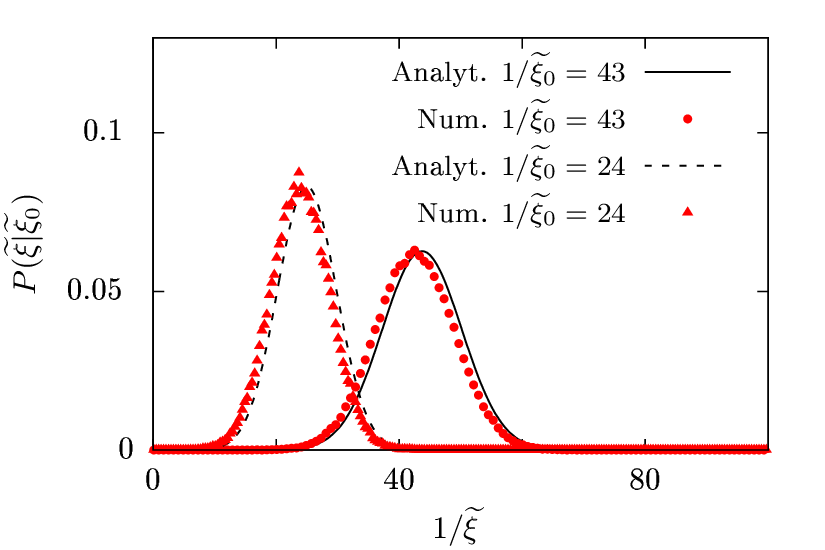}
\end{center}
\caption{
Probability distribution function $P(\widetilde{\xi}|\widetilde{\xi}_0)$ of the reduced localization length, $\widetilde{\xi}=\xi/L$, calculated analytically (Analyt.) from Eq. \ref{eq:P_gamma} and numerically (Num.) from the Anderson's model with $\varepsilon/V=-1.9$, $L=10^4 a$, and $\Delta E/V$ equal to $0.15$ and $0.20$, for $1/\xi_0=24$ and $1/\xi_0=43$ respectively.
}
\label{fig:P_xi}
\end{figure}

In Fig. \ref{fig:P_xi}, we compare the probability distribution function given by  eqs. \ref{eq:P_gamma} and  \ref{eq:xi_0} with that obtained from the histograms of numerical calculations. 
We show $P(\widetilde{\xi}|\widetilde{\xi}_0)$ for two different disorder strengths,
$\Delta E/V=0.15$ and $\Delta E/V=0.2$, which gives $1/\widetilde{\xi}_0=24$ and $1/\widetilde{\xi}_0=43$
respectively. Only small deviations were found for the conditions explored.
The value of $\xi$ for each numerical calculation with random site energies, was obtained from the direct inversion of Eq. \ref{eq:transmittance}, $1/\xi=-\lim_{L\rightarrow\infty}\ln \left( T/(2L) \right)$.
The transmittances were calculated by using the Fisher and Lee formula and the Green's functions were evaluated from the effective tight-binding Hamiltonian shown in Eq. \ref{eq:H_eff}. See Refs. \onlinecite{FPB17,PM01} for more details about this type of calculations.

Results shown above confirm that we can describe the probability distribution function of the localization length by a closed formula. Given $P(\widetilde{\xi}|\widetilde{\xi}_0)$, it is not difficult to obtain the probability distribution function of the transmittance $P(T|\widetilde{\xi}_0)$ by using Eq. \ref{eq:transmittance} and resorting to the transformation of stochastic variables,  $P(T|\widetilde{\xi}_0)= P(\widetilde{\xi}|\widetilde{\xi}_0) |d\widetilde{\xi}/dT|$.\cite{vanKampen}
The connection between $T$ and $F$ is given by Eq. \ref{eq:fuerza_lineal}. Thus, it should also be easy to obtain the probability distribution function of the nonequilibrium CIFs $P(F)$. However, as we are interested in the regime where $L \gg \xi$, it is expected that the CIF is always very close to its maximum value.

\begin{figure}[ptb]
\begin{center}
\includegraphics[width=0.45\textwidth]{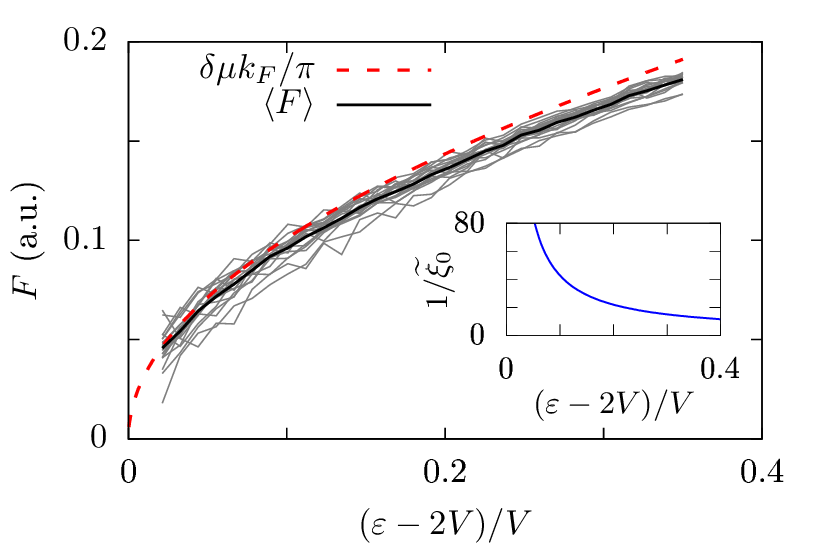}
\end{center}
\caption{
CIFs in arbitrary units, $F$, calculated numerical for the Anderson's model, Eqs. \ref{eq:Hs}-\ref{eq:Sigma}, as function of the Fermi energy $\varepsilon$. The value $(\varepsilon-2V)/V=0$ corresponds to the lower band-edge given by Eq. \ref{eq:Sigma} (close to this region the system is expected to behave as a continuum).
The Red dashed line corresponds to the theoretical maximum value of nonequilibrium CIFs, $F=\delta \mu k_F/\pi$.
Grey solid lines show the force for different disorder realizations.
The Black solid line gives the average of the grey solid lines for $N=20$ realizations.
In the calculations we used $L=10^4 a$ and $\Delta E/V=0.2$.
}
\label{fig:fuerza_vs_epsi}
\end{figure}
In Fig. \ref{fig:fuerza_vs_epsi} we compare the maximum value of the nonequilibrium CIF (red dashed line), with $T=0$, and that obtained from the tight-binding calculations.
The numerical values of the CIFs were obtained by using Eq. \ref{eq:force_scatt} on the geometry given by the panel (a) of Fig. \ref{fig:potential}.
The systems consisted of 10000 sites with random values of sites energies but with a linear smoothing over the first and last 50 sites. The smoothing function, which always makes zero the energy of the first and last sites of the chain, was used to emulate the effect of the cable coming into contact with the rotor, see appendix \ref{App:potential}. Periodic conditions in the sites energies were imposed for the explicit calculation of the derivatives of the scattering matrix, 
$\left [ \partial_x S_{\alpha,\beta} \approx S_{\alpha,\beta}(x+\delta x)-S_{\alpha,\beta}(x-\delta x)
\right ]  / \left ( 2 \delta x \right )$. 
The scattering matrices were obtained from the Green functions of the tight-binding Hamiltonian as shown in Refs. \onlinecite{PM01,FBP15,FPB17,bustos2018}.

As can be seen in Fig. \ref{fig:fuerza_vs_epsi},
Eq. \ref{eq:fuerza_lineal} (with $T \approx 0$) is in excellent agreement with the numerical simulations, especially for energies close to the band-edge. This confirms the validity of our model.
This figure also shows that  there is no real need for a $P(F|\widetilde{\xi}_0)$ under the conditions of interest, $\widetilde{\xi}_0 \ll 1$, where most of the realizations give $\xi \ll L$.
Note that at low energies there are some points where the tight-binding CIFs are
larger than the theoretical maximum value. This is only due to the fact that in
deriving Eq. \ref{fig:fuerza_vs_epsi} we neglected the reflections due to the edge
of the system. These reflections cause the appearance of small equilibrium forces
that will contribute to the total force calculated numerically.\cite{FPB17} 
Just for completeness, we give the formula for the probability distribution function of the CIFs.
\begin{equation}
  P(\widetilde{F}|\widetilde{\xi}_0)=P(\widetilde{\xi}(\widetilde{F})|\widetilde{\xi}_0)\cdot \frac{\widetilde{\xi}^2(\widetilde{F})}{1-\widetilde{F}}, \label{eq:P_F}
 \end{equation}
where we have defined the reduced CIF $\widetilde{F}=F/\left ( \delta \mu_L k_F / 2 \pi\right)$, and
$\widetilde{\xi}(\widetilde{F})=-1/\ln \left ( 1-\widetilde{F}\right )$.

The fact that nonequilibrium CIFs can be well approximated by its maximum value suggests that the efficiency $\eta$ of the
nanoelectromechanical devices build from them will always be maximum, i.e. $\eta\approx1$. However, this naive approach fails when we include in the
analysis the energy dissipated by the friction and the period of the movement. In the next section, we will see that even when we approximate $F$ by 
its maximum value, the probability distribution function of $P(\eta)$ shows a nontrivial dependence on the system parameters and the type of devices one is dealing with.

\section{Efficiency.\label{sec:efficiency}}

The performance of a nanomachine can be evaluated through its thermodynamic efficiency $\eta=P^{out}/P^{in}$, given by the ratio between the output power $P^{out}$ and the total incoming power $P^{in}$.
The former, $P^{out}$, is the difference between the power produced by the CIFs, $W/\tau$, and the power dissipated by friction, $\int_0^\tau \gamma \dot x^2 \mathrm{d}t/\tau$, where $W=\int F \mathrm{d}x$, $\gamma$ is the friction coefficient, and $\tau$ is the period of the rotor or the time during which the shuttle is being moved.
The incoming power $P^{in}$ is the current times the voltage, but the current has two contributions, the bias-dependent current $I^{\mathrm{bias}}$ and the pumped current $I^{\mathrm{pump}}$. 
At low voltages and temperatures, the bias-dependent current is given by $I^{\mathrm{bias}}(x)=(e/h)T(x) \delta \mu_{L}$, while the pumped current is given by $I^{\mathrm{pump}}=Q/\tau$.
Then, the efficiency can be written as
\begin{equation}
 \eta=\frac{W/\tau-\int_0^\tau \gamma \dot x ^2 \mathrm{d}t/\tau}{W/\tau+\frac{\left < T \right >_t}{h} \delta\mu^2 }, \label{eq:efficiency1}
\end{equation}
where we have used Eq. \ref{eq:work_Onsager}, $F^{eq}\approx 0$, and $\left < T \right >_t = \int_0^\tau T dt/\tau$.
The efficiency depends on the dynamics of the movable piece.
However, as we will show in the next subsections, it is possible to obtain closed formulas for the probability distribution function of the maximum (or optimal) efficiency reached with a given set of parameters.

In particular, we will assume negligible equilibrium forces, a constant friction coefficient, and insignificant stochastic forces. The latter implies small temperatures and/or large masses of the rotor or the shuttle, see Ref. \onlinecite{FPB17}.  
Let us take the CIFs as constant in a limit of small temperatures and voltages.
Then, the equation of motion of the system during a given time interval $t\in [0,\tau]$ is 
\begin{equation}
 F^{\mathrm{total}}=m \ddot{x}-\gamma \dot x, \label{eq:dyn_shuttle}
\end{equation}
where $m$ is the mass of the shuttle, $\dot x$ is the velocity, and $\ddot x$ the acceleration (for the rotor just replace the mass by a moment of inertia and the force by a torque).
The total force is $F^{\mathrm{total}}=F-F^{\mathrm{load}}$, i.e., the difference between the CIF $F$ and the force produced by a load
$F^{\mathrm{load}}$. This $F^{\mathrm{load}}$ can have different functional forms. For example, it can be proportional to a velocity, in which case it can be assimilated within an effective friction coefficient.
But it can also be the force needed to break or form a molecular bond, or moved the system against an electric field.
In any case, it can be treated as a correction to an effective voltage bias,
possibly an $x$-dependent one.\cite{FPB17} It all depends on the case being studied.
For the present purposes we will consider $F^{\mathrm{load}}=0$, i.e. its effect is already included in $\gamma \dot x$, or in an effective voltage bias.
Under these conditions, Eq. \ref{eq:dyn_shuttle} results in a simple first order differential equation yielding
\begin{equation}
 \dot x(t)=\frac{F}{\gamma}\left(1- e^{-\frac{\gamma}{m}t} \right), \label{eq:dynamics}
\end{equation}
where we have set $\dot x(0)=0$ and $x(0)=0$ for convenience. 

In the following subsections, we will study two different dynamical regimes, the short-time and the steady-state regimes.
For an adiabatic quantum motor like that shown in Fig. \ref{fig:potential}-(a), 
one is usually interested in steady-state conditions where the energy dissipated 
by the friction and the load is exactly compensated by the input energy and then 
$x(t)=x(t+\tau)$. On the other hand, for the shuttle shown in Fig. \ref{fig:potential}-(b), the system is expected to be far from the mentioned compensation condition and one is interested in the
short-time regime.
In this sense the examples shown in Fig. \ref{fig:potential} are complementaries.

\subsection{The shuttle in the short-time regime.\label{sec:efficiency_transient}}
 
Unlike the case of the rotor, the movement of the shuttle is not cyclic but is being driven through a linear and finite region, see scheme shown in Fig. \ref{fig:potential}-b)
For simplicity, we will assume that equilibrium forces are negligible and the friction coefficient is constant. The efficiency will depend on the specific movement followed by the shuttle.
However, just to gain some insight into its physics we will consider that the shuttle is at rest at $t=0$. Then, a constant voltage bias $\delta \mu/e$ is applied during a time $\tau$, producing a constant CIF, $F$. Under these conditions, the dynamics of the shuttle can be described by Eq. \ref{eq:dynamics}.

The efficiency is given by Eq. \ref{eq:efficiency1}, where the total work of the CIFs is now $W=F x(\tau)$.
If the total length of the guide along which the shuttle is being moved is small and the friction coefficient is also small, the short time regime holds, i.e., $\tau\ll m/\gamma$. 
Then, the movement of the shuttle can be described by $x(t) \approx F/(2m)t^2$ and the power dissipated by friction results in $\int_0^\tau \gamma \dot x ^2 \mathrm{d}t/\tau \approx \frac{2}{3}\frac{\gamma}{m} W$.
Using this, defining the dimensionless time $\widetilde{\tau}=\gamma \tau / m$, and using Eq. \ref{eq:work_Onsager} to write $(Q/e)\delta\mu/\tau=W/\tau$, where the CIFs are given by  Eq. \ref{eq:fuerza_lineal}, one can write the efficiency in the limit $\tau\ll m/\gamma$ as
\begin{equation}
 \eta=\frac{1-\frac{2}{3}\widetilde{\tau}}{1+\frac{2}{\nu \widetilde{\tau}} }.
 \label{eq:eta_shuttle1}
\end{equation}
Here, we have defined an additional dimensionless quantity, $\nu=\frac{Q^2}{x^2(\tau) \gamma T}\frac{h}{e^2}$. 
Noticing that the $x(\tau)$ factor should be used instead of $L$ in Eq. \ref{eq:pumped_charge} and that $T$ is independent of $x$, allow us to write $\nu$ as
\begin{equation}
\nu=\frac{h}{\gamma}\frac{ (1- T )^2}{T  }\left( \frac{k_F}{\pi} \right)^2. \label{eq:nu}
\end{equation}

For a given disorder and Fermi energy, the value of $\nu$ is fixed, as it depends only on intrinsic properties of the system.
However, it could still be theoretically possible to control variables such as the voltage bias or the load to manipulate $\tau$. Then, it is interesting to study the maximum value of $\eta$ accessible within a given device.
One can check that, according to Eq. \ref{eq:eta_shuttle1}, the value of $\widetilde{\tau}$ that maximizes $\eta$ is
\begin{equation}
\tau_{\mathrm{op}}=(1/\nu)\cdot \left[ -2 +  \sqrt{3\nu+4} \right], \label{eq:tau_optimo}
\end{equation}
where the subscript ``$\mathrm{op}$'' stands for ``optimal''.
Inserting Eq. \ref{eq:tau_optimo} into Eq. \ref{eq:eta_shuttle1} gives
\begin{equation}
 \eta_{\mathrm{op}}=\frac{1-\frac{2}{3\nu}\left( -2 +  \sqrt{3\nu+4} \right)}
 {1+\frac{2}{  -2 +  \sqrt{3\nu+4} } },
 \label{eq:eta_shuttle2}
\end{equation}
which is the maximum value of the efficiency in a device characterized by a given $\nu$ in the limit $\tau\ll m/\gamma$.

As shown by Eqs. \ref{eq:nu} and \ref{eq:eta_shuttle2}, $\eta$ not only depends on $T$ but on a combination of factors, given by $\nu$. 
One consequence is that having a small value of $T$ which ensures  $F \approx k_F \delta \mu_L / \pi$, not necessarily implies values of $\eta$ close to one.
The difference in the behavior of $\eta$ and $F$ for small $T$ is worsened by the nonlinear dependence of $\eta$ on the transmittance.
Therefore, it is possible to have an ensemble of nanomotors where almost all of them present CIFs close to the maximum value, but with low efficiencies.
This is why the probability distribution function of $\eta$  may be relevant even when most of the nanomotors have $\xi \ll L$.

We can obtain the probability distribution function of the optimal efficiency $P(\eta_{\mathrm{op}}|\nu_0)$ by resorting to the stochastic-variables transformation theorem,\cite{vanKampen}
\begin{equation}
P(\eta_{\mathrm{op}}|\nu_0) = P(\widetilde{\xi}(\eta_{\mathrm{op}})|\widetilde{\xi}_0(\nu_0)) 
\left|
\frac{\mathrm{d}\widetilde{\xi}}{\mathrm{d}T}
\cdot 
\frac{\mathrm{d}T}{\mathrm{d}\nu} 
\cdot 
\frac{\mathrm{d}\nu}{\mathrm{d}\eta_{\mathrm{op}}} 
\right|. \label{eq:Probab_shuttle}
\end{equation}
The first two derivatives in Eq. \ref{eq:Probab_shuttle} can be obtained from 
Eqs. \ref{eq:transmittance} and \ref{eq:nu}, giving
\begin{equation}
 \frac{\mathrm{d}\widetilde{\xi}}{\mathrm{d}T} = \frac{\widetilde{\xi}^2 }{2T} 
 \mbox{\quad and \quad} 
 \frac{\mathrm{d}T}{\mathrm{d}\nu} = -\frac{T }{\nu}. \label{eq:dxidT-dTdnu} 
\end{equation}
Note that we used $1-T \approx 1$ for the second inverse function as we are interested in the regime $\widetilde{\xi} \ll 1$.
To obtain $\frac{\mathrm{d}\nu}{\mathrm{d}\eta_{\mathrm{op}}} $ we need the inverse of Eq. \ref{eq:eta_shuttle2}, which is
\begin{equation}
\nu=\frac{16}{3}\frac{\eta_{\mathrm{op}}}{(1-\eta_{\mathrm{op}})^2}. \label{eq:nu(eta_op)}
\end{equation}
Then, the last derivative needed in Eq. \ref{eq:Probab_shuttle} is
\begin{equation}
\frac{\mathrm{d}\nu}{\mathrm{d}\eta_{\mathrm{op}}} = \frac{\nu}{\eta_{\mathrm{op}}} \cdot \frac{1+\eta_{\mathrm{op}}}{(1-\eta_{\mathrm{op}})}. \label{eq:dnudeta}
 \end{equation}
As shown by Eq. \ref{eq:P_F}, once one works with the reduced CIF, $\widetilde{F}$, the probability distribution function of the CIFs is controlled by only one parameter, $\widetilde{\xi_0}$.
Regretfully, the probability distribution function of $\eta_{\mathrm{op}}$ cannot be written only in terms of $\widetilde{\xi_0}$, as it truly depends on other parameters, namely the friction coefficient $\gamma$ and $k_F$.
However, this issue can be solved by defining $T_0$ as the value of $T$ obtained by replacing $\xi$ by $\xi_0$ in Eq. \ref{eq:transmittance}, and then $\nu_0$ as the value of $\nu$ obtained by replacing $T$ by $T_0$ in Eq. \ref{eq:nu} (with $k_F$ and $\gamma$ fixed).
Then, one can use $\nu_0$ as the single parameter that controls the probability distribution function of $\eta_{\mathrm{op}}$, $P(\eta_{\mathrm{op}}|\nu_0)$.
Using Eqs. \ref{eq:Probab_shuttle}-\ref{eq:dnudeta} one obtains
\begin{eqnarray}
P(\eta_{\mathrm{op}}|\nu_0) &=& P(\widetilde{\xi}|\widetilde{\xi}_0) 
\left| \frac{\mathrm{d}\widetilde{\xi}}{\mathrm{d}\eta_{\mathrm{op}}} \right| , \nonumber \\
 &=& P(\widetilde{\xi}|\widetilde{\xi}_0) 
 \left( \frac{\widetilde{\xi}^2  }{2} \frac{1+\eta_{\mathrm{op}}}{\eta_{\mathrm{op}}(1-\eta_{\mathrm{op}})} \right). \label{eq:P_eta_AQM}
\end{eqnarray}
where $\widetilde{\xi} \equiv \widetilde{\xi}(\eta_{\mathrm{op}})$ and $\widetilde{\xi}_0 \equiv \widetilde{\xi}_0(\nu_0)$. The explicit dependence of $\widetilde{\xi}$ with $\eta_{\mathrm{op}}$ can be obtained by combining Eqs. \ref{eq:nu(eta_op)}, \ref{eq:nu} (in the limit $T \ll 1$), and \ref{eq:transmittance},
\begin{equation}
\widetilde{\xi}= 2 \left[ \ln \left( \frac{16 \gamma \pi^2}{3 h k_F^2}
\frac{\eta_{\mathrm{op}}}{(1-\eta_{\mathrm{op}})^2} \right) \right]^{-1}.
\label{eq:xi(eta)}
\end{equation}

\begin{figure}[ptb]
\begin{center}
    \includegraphics[width=0.5\textwidth]{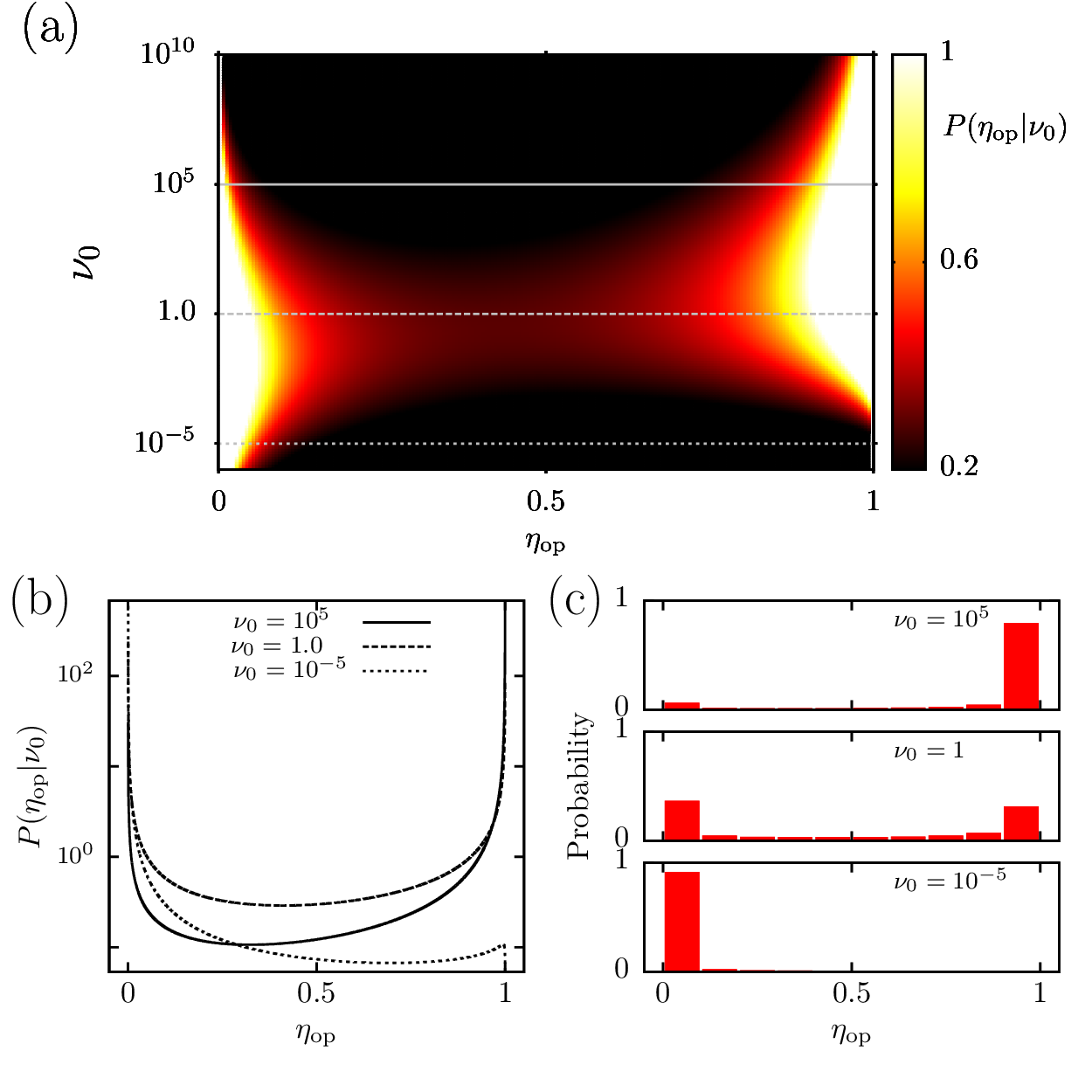}
\end{center}
\caption{
$(a)$ Probability distribution function of the optimal efficiency $P(\eta_{\mathrm{op}}|\nu_0)$ as function of the optimal efficiency $\eta_{\mathrm{op}}$ and the parameter $\nu_0$ (see discussion above Eq. \ref{eq:xi(eta)}).
 $(b)$ Horizontal cuts of panel $a$. $(c)$ Probability of achieving a given interval of optimum efficiency, between 
 $\eta_{\mathrm{op}}$ and $\eta_{\mathrm{op}}+\delta \eta_{\mathrm{op}}$ ($\delta \eta_{\mathrm{op}}=0.1$).
 }
\label{fig:P_eta_map}
\end{figure}

In Fig. \ref{fig:P_eta_map}-$(a)$ we plot the probability distribution function of the optimal efficiency $P(\eta_{\mathrm{op}}|\nu_0)$ (in colors), as function of $\eta_{\mathrm{op}}$ and $\nu_0$. 
We can see that the most probable efficiencies gather around $\eta_{\mathrm{op}}=0$ and $\eta_{\mathrm{op}}=1$ with a clear dependence on $\nu_0$.
For small values of $\nu_0$ ($\nu_0 \ll 1$), almost all nanomotors are inefficients, while for large ones ($\nu_0 \gg 1$), almost all nanomotors are highly efficient. This can be better appreciated in Figs. \ref{fig:P_eta_map}-$(b)$ and \ref{fig:P_eta_map}-$(c)$.
There is no clear cut between these two regimes but according to Fig. \ref{fig:P_eta_map} the region of intermediate behavior is around $\nu_0 \approx 1$.  Equations \ref{eq:transmittance}, \ref{eq:xi_0}, and \ref{eq:nu}, together with $\nu_0 > 1$, allow us to write the condition for the minimum disorder strength needed to ensure efficient nanomotors:
\begin{equation}
\Delta E^2 a > 12 \frac{v_F^2 \hbar^2}{L} \ln{\left [ \frac{\gamma }{h} \left( \frac{\pi}{k_F} \right)^2 \right ]}. \label{eq:feasibility}
\end{equation}
In appendix \ref{App:estimation} we discuss the feasibility of AAQMs based on the above equation.

\subsection{The rotor in the steady-state regime.\label{sec:efficiency_Steady}}

Once the rotor reaches the steady-state regime, the terminal velocity $\dot x$ can be approximated by $L/\tau $, where $L$ is the rotor's perimeter here. Then, the power dissipated by friction becomes $\int_0^\tau \gamma \dot x ^2 \mathrm{d}t/\tau \simeq \left < \gamma \right > L^2/\tau^2$,  where $\left < \gamma \right > = \int_0^\tau \gamma \mathrm{d}t/\tau$.
As discussed in Refs. \onlinecite{FPB17,calvo2017}, the above approximations will be accurate when the average kinetic energy of the system at steady-state is much larger than the difference between the maximum and minimum of the potential energy given by the equilibrium forces.
This case holds for large moments of inertia, large voltages, or small friction coefficients. Stochastic forces can also take us away from the approximation $\dot x \approx L/\tau$ but their effect on the dynamics diminishes when the moment of inertia increases or when the temperature decreases.
In summary, the expression we are about to discuss should be accurate under the former conditions, see Refs. \onlinecite{FPB17,calvo2017}, but one should keep in mind that some deviation may appear for realistic systems, especially for small terminal velocities (large $\tau$'s) where equilibrium forces can dramatically alter the dynamics leading to hysteresis-like cycles for example.\cite{FPB17,calvo2017}
One should also keep in mind that, high terminal velocities, small $\tau$'s, could break the adiabatic approximation leading to deviations of the equations of motion. \cite{bode2012,FPB17}

We start by rewriting Eq. \ref{eq:efficiency1} using Eq. \ref{eq:work_Onsager} to write $(Q/e)\delta\mu/\tau=W/\tau$, and defining the dimensionless period $\widetilde{\tau}= \frac{(Q/e)\delta\mu}{L^2 \gamma} \tau$. This yields
\begin{equation}
 \eta=\frac{1- \frac{1}{\widetilde{\tau}} }{1+\frac{\widetilde{\tau}}{\nu} }. \label{eq:efficiency2}
\end{equation}
As before, we have defined the dimensionless quantity
$\nu=\frac{ Q^2}{L^2 \gamma \left< T\right>_t}\frac{h}{e^2}$, which can be simplified using Eq. \ref{eq:pumped_charge},
\begin{equation}
\nu=\frac{h}{\gamma}\frac{ (1-\left<T\right>_x)^2}{\left<T\right>_t }\left( \frac{k_F}{\pi} \right)^2. \label{eq:nu2}
\end{equation}

The optimal value of $\widetilde{\tau}$ that maximizes the efficiency is given by
\begin{equation}
\widetilde{\tau}_{\mathrm{op}}=1+\sqrt{1+\nu}.
\end{equation}
Now evaluating $\eta$ at $\widetilde{\tau}_{\mathrm{op}}$ yields
\begin{equation}
\eta_{\mathrm{op}}=\frac{2+\nu-2\sqrt{1+\nu}}{\nu}.
\label{eq:eta_op}
\end{equation}

To obtain closed formulas for $P(\eta_{\mathrm{op}}|\nu_0)$, one requires the expression for $P(\left<T\right>_x)$, the probability distribution function of the transmittances averaged over a full cycle of the rotor. We numerically study $P(\left<T\right>_x)$ and found two limiting situations where it can be easily calculated, see appendix \ref{App:PT}. When the wire completely wraps the rotor, $P(\left<T\right>_x)$ can be approximated by  $P(T)$.
This finding is, at present, based only on numerical evidence for the used parameters, $\Delta E \ll V$ (the weak disorder limit), the Fermi energy close to the band edge, and $L/a \gg 1$ (such that $T \ll 1$).
In the opposite case, a small contact region, $P(\left < T \right >_x)$ becomes a narrow function centered around $\left < T \right >$, the average value of $T$ over different sampling of impurities. There, $T$ can be taken as constant, as well as $\eta$.
The explanation for this latter limiting situation is simple.
In the limit of $R \rightarrow \infty$ at $L$ constant a full rotation of the systems implies that every possible combination of impurities have been sampled for $T$, then $\left < T \right >_x$ is simply $\left < T \right >$.
For intermediate regimes, the function $P(\left < T \right >_x)$ is more difficult to model providing a smooth transition between the two other regimes.
However, once obtained, numerically, for example, the formulas we are about to discuss can be straightforwardly corrected following the same procedure to the one discussed here and in the previous section.
Finally, one last assumption has to be made in Eq. \ref{eq:nu2}, $\left<T\right>_x \approx \left<T\right>_t$. 
In this case, averaging over time or averaging over the coordinate are the same for a rotor moving at a constant velocity, approximation discussed at the beginning of this section.

Taking into account the above discussion, we propose the following concrete functional form for $P ( \left < T\right >_x)$, which should be a good approximation, according to numerical evidence, when the wire completely wraps the rotor,
\begin{equation}
 P ( \left < T\right >_x | \xi_0) \approx P(\widetilde{\xi} | \xi_0)
 \frac{\mathrm{d} \widetilde{\xi} }{\mathrm{d} \left < T\right >_x }. \label{eq:PT_x}
\end{equation}
Here, $P(\widetilde{\xi} | \xi_0)$ is given by Eq. \ref{eq:P_gamma} and $\left < T\right >_x = \exp[-2/\widetilde{\xi}]$.
With this functional form, the equality $P(\left<T\right >_x|\xi_0)=P(T|\xi_0)$ is obviously fulfilled.

To calculate the distribution function $P(\eta_{\mathrm{op}}|\nu_0)$, we need the following inverse functions \begin{eqnarray}
\widetilde{\xi} &=& -2\left ( \ln \left<T\right >_x \right )^{-1}
, \notag \\ 
\mbox{\quad}
\left<T\right >_x & \approx & \frac{h}{\gamma \nu} \left( \frac{k_F}{\pi} \right)^2
, \mbox{\quad and \quad} \notag \\ 
 \nu &=& \frac{4\eta_{\mathrm{op}}}{(1-\eta_{\mathrm{op}})^2}.
\label{eq:inverses_SS}
\end{eqnarray}
Note that the last formula is proportional to $\nu(\eta_{\mathrm{op}})$ for the short-time regime given in Eq. \ref{eq:nu(eta_op)} (with a factor $4$ instead of $\frac{16}{3}$). Combining the above expressions allow us to obtain the relation between $\widetilde{\xi}$ and $\eta_{\mathrm{op}}$
\begin{equation}
\widetilde{\xi}= 2 \left[ \ln \left( \frac{4 \gamma \pi^2}{h k_F^2}
\frac{\eta_{\mathrm{op}}}{(1-\eta_{\mathrm{op}})^2} \right) \right]^{-1}.
\label{eq:xi(eta2)}
\end{equation}
Using the above, one can find that the approximated probability distribution function of the maximum efficiency of the example treated here results in exactly the same as that shown in Eq. \ref{eq:P_eta_AQM}.
Therefore, the discussion about Fig. \ref{fig:P_eta_map} remains the same for the present case, as well as the condition for the minimum disorder strength needed to ensure efficient nanomotors, see Eq. \ref{eq:feasibility} and the discussion in appendix \ref{App:estimation}.

\section{Conclusions.\label{sec:Conclusions}}

We have proposed what we called an Anderson adiabatic quantum motor (AAQM), i.e, a current-driven nanomotor based on Anderson's localization.
We have studied two geometries for AAQMs, the shuttle and the rotor (see Fig. \ref{fig:potential}). We have derived general expressions to evaluate the nonequilibrium current-induced forces (Eq. \ref{eq:force}) as well as the efficiency (Eq. \ref{eq:efficiency1}) of this kind of devices.
Due to the stochastic nature of AAQMs, we based our analysis on the probability distribution functions of the properties of interest.
We have shown that, under a certain regime of parameters, most of the disorder realizations result in systems with a maximal value of the current-induced forces, where the reflectance is almost one.
However, the same regime of parameters not necessarily leads to a maximum efficiency.
We have studied the performance of these devices in the short-time dynamical regime and under steady-state conditions.
We have found an analytical expression of the probability distribution function of the maximum efficiency of the shuttle, see Eqs. \ref{eq:P_eta_AQM} and \ref{eq:xi(eta)}.
For the rotor, we have numerically found that, under certain conditions, the probability distribution function of the transmittances averaged over one period is well described by a simple formula that describes the probability distribution function of transmittances in the Anderson's model of disorder. Using this, we have shown that, under certain conditions, both dynamical regimes (the rotor in the steady-state regime and the shuttle in the short-time regime) present very similar probability distribution functions of their maximum efficiency (Eq. \ref{eq:P_eta_AQM}) despite having quite different expressions for their efficiencies (Eqs. \ref{eq:eta_shuttle1} and \ref{eq:efficiency2}).
Finally, we provide an expression to estimate the minimal disorder strength required to obtain efficient nanomotors (Eq. \ref{eq:feasibility}).

As compared with other proposals of adiabatic quantum motors,
\cite{AQM13,FBP15,arrachea2015,Silvestrov2016,Liliana_Onsager,FPB17,calvo2017}
the AAQMs require, in principle, less control over the impurities
or charges responsible for the position-dependent coupling between the electrons 
and the moving piece of the nanomotor. 
For this reason, we believe AAQMs should be easier to realize than other proposed adiabatic quantum motors.
One drawback, which is common to most adiabatic quantum motors, is that AAQMs would require coherence lengths of the order of the nanodevice itself. Then, it would be interesting to understand to what extent AAQMs can tolerate decoherence in relation to the amount of disorder they possess.
Although preliminary estimations seem promising, it would be important to study numerically concrete examples of AAQMs to evaluate their feasibility under realistic conditions.
From a theoretical point of view, it would be interesting to understand the reason behind the found similarity between the probability distribution function of the transmittance of the rotor at a fixed position $P\left ( T \right )$ and the probability distribution function of the transmittance averaged over one cycle $P\left ( \left <T \right >_x \right )$.
Finally, the connection between disorder-induced localization and incommensurability \cite{PWA83,WP82} may open the door to another type of closely related adiabatic quantum motor.

\section{Acknowledgements}

This work was supported by Consejo Nacional de Investigaciones Cient\'ificas y T\'ecnicas (CONICET), Argentina;
Secretar\'ia de Ciencia y Tecnolog\'ia, Universidad Nacional de C\'ordoba (SECYT-UNC);
and Ministerio de Ciencia y Tecnolog\'ia de la Provincia de C\'ordoba (MinCyT-Cor).

\appendix

\section{Estimation of disorder needed in an AAQM.\label{App:estimation}}

It is difficult to make general statements about the feasibility of AAQMs without resorting to particular cases. For example, the friction coefficient is expected to depend on the contact surface between the rotor and the wire (or the shuttle and the guide), $v_F$ depends on the material and its doping, the characteristic of the disorder and thus the parameters that describe it ($\Delta E$ and $a$) will depend on how the disorder is realized, etc.
However, just for the sake of making a rough estimation let us take a concrete example of AAQMs with $a=2nm$, $L=200nm$, $v_F=10^6 m/s$, $k_F=10^{10} m^{-1}$, and $\gamma=2.5 \times 10^{-8} kg/s$.
\footnote{
The values of $v_F$ and $k_F$ were taken from the order of magnitude of typical metals.\cite{AshcroftMermin} The value of $\gamma$ used ($2.5 \times 10^{-8} kg/s$) implies $\tau = 1 \times 10^{-2}s$ at $\delta \mu = 1 \times 10^{-3} eV$. This was estimated from $\delta \mu k_F/\pi = \gamma \dot x$ and assuming a constant terminal velocity with $F^{load} = 0$. The value of $\tau = 1 \times 10^{-2}s$ is of the order of the nanomotor reported by K. Kim et. al.\cite{FastNM}.
}
With these values and according to Eq. \ref{eq:feasibility}, the minimum disorder needed results in $\Delta E = 0.44eV$, which is about $10\%$ of the hopping parameter corresponding to a $\pi$ bond between carbon atoms in a conductive polymer.\cite{CBMP10}

As mentioned the friction coefficient may change substantially from one device to another. However, $\Delta E$ depends only logarithmically on it, so the above estimation should be robust against a variation of $\gamma$. On the other hand, $v_F$ was taken from the order of magnitude of typical metals,\cite{AshcroftMermin} where the Fermi energy is at the center of the conduction band. For energies closer to a band-edge, $v_F$ is expected to be much smaller, which should reduce considerably the minimum value of the energy uncertainty required.

\section{On the electron's potential in a wire coiled around a rotor with fixed charges.\label{App:potential}}
\begin{figure}[ptb]
\begin{center}
\includegraphics[width=2.5 in]{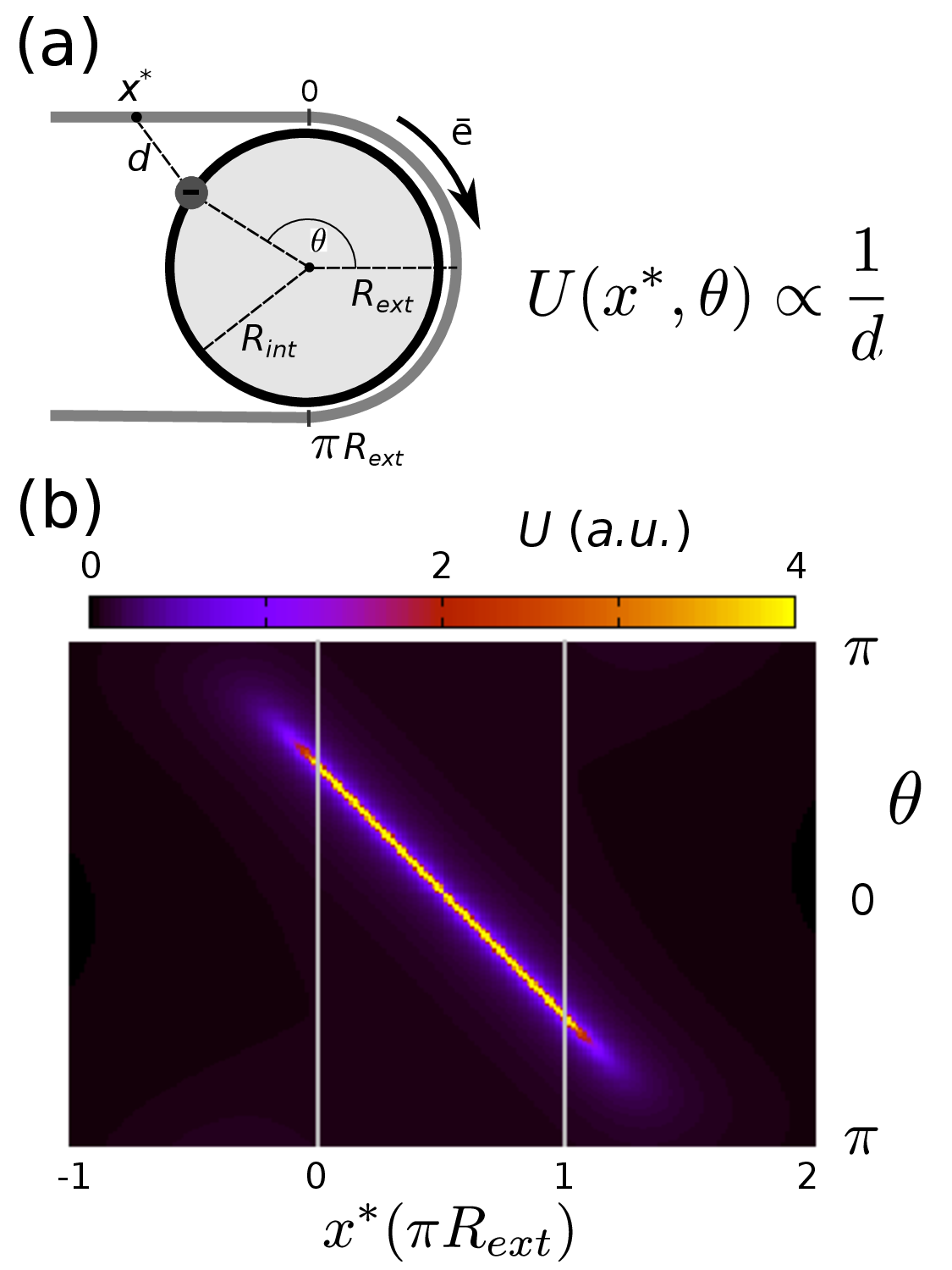}
\end{center}
\caption{
Example of the potential experienced by the electron in a wire coiled around a rotor with a fixed charge. (a) Geometry considered. $R_{int}$ is the radius of the rotor, $R_{ext}-R_{int}$ is the minimum separation between the wire and the rotor, $\theta$ is the angle that sets the position of the rotor, and $x^*$ is the coordinate along the wire. The example assumes a potential of the form $U\propto 1/d$ where $d$ is the distance between the fixed charge and the point along the wire (set by $x^*$). (b) Potential in arbitrary units as a function of the position along the wire ($x^*$) and the position of the rotor ($\theta$).
In the figure we set $R_{int}=0.99 R_{ext}$.
}
\label{fig:extra}
\end{figure}
In Fig. \ref{fig:extra} we show a simple example of the interaction between a rotor with a fixed charge and the electrons in a wire. As can be seen in panel (b), the effect of a rotation of the rotor on the potential sensed by the electrons can be modeled as a scatterer that appears from nowhere that then moves in a certain direction until it disappears again. The details of how the ``scatterer'' appears and disappears depend, of course, on the details of how the wire is coiled around the rotor. However, the shift of the ``scatterer'' with $\theta$ in a certain region is a universal characteristic that is just consequence of the fixed distance between the wire and the rotor in that region.
For more complex potentials, caused by random charges, for example, the effect of a rotation of $\theta$ is the same. There is a small region from where new features of the potential gently appear, a region where there is a shift of the potential with $\theta$, and a small region where the features of the potential gently disappear.
In the numerical simulations discussed around Fig. \ref{fig:fuerza_vs_epsi}, we modeled the dependence of the potential with $\theta$ in precisely that way. We tried different smoothing function (linear and Gaussian) to describe the appearance and disappearance of potential's features, but only a small effect on the equilibrium part of the forces was observed. The same behavior was observed in the Thouless motor studied previously.\cite{FPB17}

The simple example analyzed here illustrates the mechanism behind the adiabatic quantum motors studied in this work, and the related adiabatic quantum pumps. They are caused by the ``snow-plow'' effect\cite{Avron2004,Cohen2005,Nazarov2009} and momentum conservation of the reflected electrons.
Describing the movement of the rotor by the Cartesian coordinates of a point over its surface, one can readily check that the trajectories will enclose the origin. This implies a net shift of the ``scatterers'' as in the case of impurities, along a conductor, being moved by the current, see for example section 1.7.4 of Ref. \onlinecite{Nazarov2009}. The difference is that here the features of the potential (or ``scatterers'') appear from nowhere in a region and disappear in another region.
A classical picture that can also help to understand the mechanism behind the rotor shown in panel (a) of Fig. \ref{fig:potential}, is that of a water wheel but with paddles randomly placed. The difference with this classical analog is that the potential energy caused by the ``paddles'' is smaller than the kinetic energy of the electrons. Thus, only quantum interferences can explain the reflection of the electrons and the movement of the rotor.
\newline

\section{Differences between $P\left ( \left< T \right>_x \right )$ and $P(T)$ \label{App:PT}}
\begin{figure}
\begin{center}
\includegraphics[width=3.3 in]{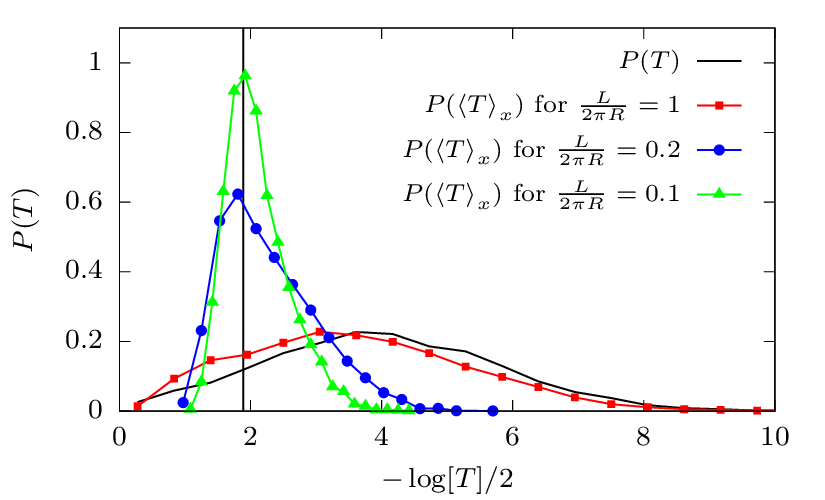}
\end{center}
\caption{
Comparison between $P\left ( \left< T \right>_x \right )$ and $P(T)$. In the numerical simulations we used $\Delta E/V = 0.2$, $L=10^3 a$, and $\varepsilon/V=1.9$. The vertical line is $-\log[\left < T \right >]/2$, where $\left < T \right >$ is the average value of $T$.
}
\label{fig:extra2}
\end{figure}
To account for the differences between the probability distribution function of $\left < T \right >_x$ and $T$, $P(\left < T \right >_x)$ and $P(T)$ respectively, we performed a set of numerical calculations using the same tight-binding Hamiltonian as that shown in sec. \ref{sec:CIFs_Anderson}.
In our calculations, we first sampled $N_r$ site's energies using a uniform probability distribution function of width $\Delta E$.
The Hamiltonian of the system was then constructed with the consecutive site's energies $E_j$ starting from $j=j_0$ and finishing with $j=j_0+N_{sys}$, with the periodic condition $E_j=E_{j+N_r}$.
The number of sites of the system was kept fixed in the simulations, $N_{sys}=1000$. As in section \ref{sec:CIFs_Anderson}, we  imposed a linear smoothing over the first and last 50 sites and added a self-energy to first and last sites of the system.
The transmittances were obtained from the Green function of the tight-binding Hamiltonian as shown in Refs. \onlinecite{PM01,FBP15,FPB17,bustos2018}.
All this was done to emulate a rotor with $N_r$ sites ($2 \pi R =N_{r} a$) in contact with a wire, where the contact region involved $N_{sys}$ sites ($L=N_{sys} a$).
The value of $\left < T \right >_x$ was obtained by averaging $T$ over a cycle of the rotor, $j_0$ from $1$ to $N_r$.
We repeated this procedure to obtain  a set of $\left < T \right >_x$ values and made a normalized histrogram to obtain $P(\left < T \right >_x)$. $P(T)$ was obtained from the same simulations but with fixed $j_0$. 

Some representative results of our calculations are shown in Fig. \ref{fig:extra2}.
There, one can notice that the behavior of $P(\left < T \right >_x)$ depends strongly on the ratio $L/2 \pi R$, the ratio between the contact region ($L$) and the perimeter of the rotor ($2 \pi R$).
However, two important limiting situations can be distinguished. 
When the wire completely wraps the rotor, $P(\left < T \right >_x)$ and $P(T)$ are very similar. In the opposite limit, when the wire barely touch the rotor, $P(\left < T \right >_x)$ becomes a narrow function centered around the average value of $T$,  $\left < T \right > = \int T P(T) dT$.

\bibliography{./references-v7p1}

\begin{thebibliography}{56}%
\makeatletter
\providecommand \@ifxundefined [1]{%
 \@ifx{#1\undefined}
}%
\providecommand \@ifnum [1]{%
 \ifnum #1\expandafter \@firstoftwo
 \else \expandafter \@secondoftwo
 \fi
}%
\providecommand \@ifx [1]{%
 \ifx #1\expandafter \@firstoftwo
 \else \expandafter \@secondoftwo
 \fi
}%
\providecommand \natexlab [1]{#1}%
\providecommand \enquote  [1]{``#1''}%
\providecommand \bibnamefont  [1]{#1}%
\providecommand \bibfnamefont [1]{#1}%
\providecommand \citenamefont [1]{#1}%
\providecommand \href@noop [0]{\@secondoftwo}%
\providecommand \href [0]{\begingroup \@sanitize@url \@href}%
\providecommand \@href[1]{\@@startlink{#1}\@@href}%
\providecommand \@@href[1]{\endgroup#1\@@endlink}%
\providecommand \@sanitize@url [0]{\catcode `\\12\catcode `\$12\catcode
  `\&12\catcode `\#12\catcode `\^12\catcode `\_12\catcode `\%12\relax}%
\providecommand \@@startlink[1]{}%
\providecommand \@@endlink[0]{}%
\providecommand \url  [0]{\begingroup\@sanitize@url \@url }%
\providecommand \@url [1]{\endgroup\@href {#1}{\urlprefix }}%
\providecommand \urlprefix  [0]{URL }%
\providecommand \Eprint [0]{\href }%
\providecommand \doibase [0]{http://dx.doi.org/}%
\providecommand \selectlanguage [0]{\@gobble}%
\providecommand \bibinfo  [0]{\@secondoftwo}%
\providecommand \bibfield  [0]{\@secondoftwo}%
\providecommand \translation [1]{[#1]}%
\providecommand \BibitemOpen [0]{}%
\providecommand \bibitemStop [0]{}%
\providecommand \bibitemNoStop [0]{.\EOS\space}%
\providecommand \EOS [0]{\spacefactor3000\relax}%
\providecommand \BibitemShut  [1]{\csname bibitem#1\endcsname}%
\let\auto@bib@innerbib\@empty
\bibitem [{\citenamefont {Craighead}(2000)}]{Craighead2000_NEMS}%
  \BibitemOpen
  \bibfield  {author} {\bibinfo {author} {\bibfnamefont {H.~G.}\ \bibnamefont
  {Craighead}},\ }\href {\doibase 10.1126/science.290.5496.1532} {\bibfield
  {journal} {\bibinfo  {journal} {Science}\ }\textbf {\bibinfo {volume}
  {290}},\ \bibinfo {pages} {1532} (\bibinfo {year} {2000})}\BibitemShut
  {NoStop}%
\bibitem [{\citenamefont {Roukes}(2001)}]{roukes2001_NEMS}%
  \BibitemOpen
  \bibfield  {author} {\bibinfo {author} {\bibfnamefont {M.}~\bibnamefont
  {Roukes}},\ }\href@noop {} {\bibfield  {journal} {\bibinfo  {journal}
  {Physics World}\ }\textbf {\bibinfo {volume} {14}},\ \bibinfo {pages} {25}
  (\bibinfo {year} {2001})}\BibitemShut {NoStop}%
\bibitem [{\citenamefont {Michl}\ and\ \citenamefont
  {Sykes}(2009)}]{MolcecMotorsGral}%
  \BibitemOpen
  \bibfield  {author} {\bibinfo {author} {\bibfnamefont {J.}~\bibnamefont
  {Michl}}\ and\ \bibinfo {author} {\bibfnamefont {E.~C.~H.}\ \bibnamefont
  {Sykes}},\ }\href@noop {} {\bibfield  {journal} {\bibinfo  {journal} {ACS
  Nano}\ }\textbf {\bibinfo {volume} {3}},\ \bibinfo {pages} {1042} (\bibinfo
  {year} {2009})}\BibitemShut {NoStop}%
\bibitem [{\citenamefont {Dundas}\ \emph {et~al.}(2009)\citenamefont {Dundas},
  \citenamefont {McEniry},\ and\ \citenamefont {Todorov}}]{Dundas}%
  \BibitemOpen
  \bibfield  {author} {\bibinfo {author} {\bibfnamefont {D.}~\bibnamefont
  {Dundas}}, \bibinfo {author} {\bibfnamefont {E.~J.}\ \bibnamefont {McEniry}},
  \ and\ \bibinfo {author} {\bibfnamefont {T.~N.}\ \bibnamefont {Todorov}},\
  }\href@noop {} {\bibfield  {journal} {\bibinfo  {journal} {Nature Nanotech.}\
  }\textbf {\bibinfo {volume} {4}},\ \bibinfo {pages} {99} (\bibinfo {year}
  {2009})}\BibitemShut {NoStop}%
\bibitem [{\citenamefont {Bailey}\ \emph {et~al.}(2008)\citenamefont {Bailey},
  \citenamefont {Amanatidis},\ and\ \citenamefont {Lambert}}]{Bailey08}%
  \BibitemOpen
  \bibfield  {author} {\bibinfo {author} {\bibfnamefont {S.~W.~D.}\
  \bibnamefont {Bailey}}, \bibinfo {author} {\bibfnamefont {I.}~\bibnamefont
  {Amanatidis}}, \ and\ \bibinfo {author} {\bibfnamefont {C.~J.}\ \bibnamefont
  {Lambert}},\ }\href@noop {} {\bibfield  {journal} {\bibinfo  {journal} {Phys.
  Rev. Lett.}\ }\textbf {\bibinfo {volume} {100}},\ \bibinfo {pages} {256802}
  (\bibinfo {year} {2008})}\BibitemShut {NoStop}%
\bibitem [{\citenamefont {Bustos-Mar\'un}(2018)}]{bustos2018}%
  \BibitemOpen
  \bibfield  {author} {\bibinfo {author} {\bibfnamefont {R.~A.}\ \bibnamefont
  {Bustos-Mar\'un}},\ }\href {\doibase 10.1103/PhysRevB.97.075412} {\bibfield
  {journal} {\bibinfo  {journal} {Phys. Rev. B}\ }\textbf {\bibinfo {volume}
  {97}},\ \bibinfo {pages} {075412} (\bibinfo {year} {2018})}\BibitemShut
  {NoStop}%
\bibitem [{\citenamefont {McEniry}\ \emph {et~al.}(2009)\citenamefont
  {McEniry}, \citenamefont {Todorov},\ and\ \citenamefont
  {Dundas}}]{McEniry09current_cooling}%
  \BibitemOpen
  \bibfield  {author} {\bibinfo {author} {\bibfnamefont {E.~J.}\ \bibnamefont
  {McEniry}}, \bibinfo {author} {\bibfnamefont {T.~N.}\ \bibnamefont
  {Todorov}}, \ and\ \bibinfo {author} {\bibfnamefont {D.}~\bibnamefont
  {Dundas}},\ }\href@noop {} {\bibfield  {journal} {\bibinfo  {journal}
  {Journal of Physics: Condensed Matter}\ }\textbf {\bibinfo {volume} {21}},\
  \bibinfo {pages} {195304} (\bibinfo {year} {2009})}\BibitemShut {NoStop}%
\bibitem [{\citenamefont {Galperin}\ \emph {et~al.}(2009)\citenamefont
  {Galperin}, \citenamefont {Saito}, \citenamefont {Balatsky},\ and\
  \citenamefont {Nitzan}}]{galperin2009cooling}%
  \BibitemOpen
  \bibfield  {author} {\bibinfo {author} {\bibfnamefont {M.}~\bibnamefont
  {Galperin}}, \bibinfo {author} {\bibfnamefont {K.}~\bibnamefont {Saito}},
  \bibinfo {author} {\bibfnamefont {A.~V.}\ \bibnamefont {Balatsky}}, \ and\
  \bibinfo {author} {\bibfnamefont {A.}~\bibnamefont {Nitzan}},\ }\href@noop {}
  {\bibfield  {journal} {\bibinfo  {journal} {Physical Review B}\ }\textbf
  {\bibinfo {volume} {80}},\ \bibinfo {pages} {115427} (\bibinfo {year}
  {2009})}\BibitemShut {NoStop}%
\bibitem [{\citenamefont {Arrachea}\ \emph {et~al.}(2012)\citenamefont
  {Arrachea}, \citenamefont {Mucciolo}, \citenamefont {Chamon},\ and\
  \citenamefont {Capaz}}]{arrachea2012refrigerator}%
  \BibitemOpen
  \bibfield  {author} {\bibinfo {author} {\bibfnamefont {L.}~\bibnamefont
  {Arrachea}}, \bibinfo {author} {\bibfnamefont {E.~R.}\ \bibnamefont
  {Mucciolo}}, \bibinfo {author} {\bibfnamefont {C.}~\bibnamefont {Chamon}}, \
  and\ \bibinfo {author} {\bibfnamefont {R.~B.}\ \bibnamefont {Capaz}},\
  }\href@noop {} {\bibfield  {journal} {\bibinfo  {journal} {Physical Review
  B}\ }\textbf {\bibinfo {volume} {86}},\ \bibinfo {pages} {125424} (\bibinfo
  {year} {2012})}\BibitemShut {NoStop}%
\bibitem [{\citenamefont {Fern\'andez-Alc\'azar}\ \emph
  {et~al.}(2017)\citenamefont {Fern\'andez-Alc\'azar}, \citenamefont
  {Pastawski},\ and\ \citenamefont {Bustos-Mar\'un}}]{FPB17}%
  \BibitemOpen
  \bibfield  {author} {\bibinfo {author} {\bibfnamefont {L.~J.}\ \bibnamefont
  {Fern\'andez-Alc\'azar}}, \bibinfo {author} {\bibfnamefont {H.~M.}\
  \bibnamefont {Pastawski}}, \ and\ \bibinfo {author} {\bibfnamefont {R.~A.}\
  \bibnamefont {Bustos-Mar\'un}},\ }\href@noop {} {\bibfield  {journal}
  {\bibinfo  {journal} {Phys. Rev. B}\ }\textbf {\bibinfo {volume} {95}},\
  \bibinfo {pages} {155410} (\bibinfo {year} {2017})}\BibitemShut {NoStop}%
\bibitem [{\citenamefont {Kim}\ \emph {et~al.}(2014)\citenamefont {Kim},
  \citenamefont {Xu}, \citenamefont {Guo},\ and\ \citenamefont {Fan}}]{FastNM}%
  \BibitemOpen
  \bibfield  {author} {\bibinfo {author} {\bibfnamefont {K.}~\bibnamefont
  {Kim}}, \bibinfo {author} {\bibfnamefont {X.}~\bibnamefont {Xu}}, \bibinfo
  {author} {\bibfnamefont {J.}~\bibnamefont {Guo}}, \ and\ \bibinfo {author}
  {\bibfnamefont {D.~L.}\ \bibnamefont {Fan}},\ }\href@noop {} {\bibfield
  {journal} {\bibinfo  {journal} {Nat. Commun.}\ }\textbf {\bibinfo {volume}
  {5}},\ \bibinfo {pages} {3632} (\bibinfo {year} {2014})}\BibitemShut
  {NoStop}%
\bibitem [{\citenamefont {Kudernac}\ \emph {et~al.}(2011)\citenamefont
  {Kudernac}, \citenamefont {Ruangsupapichat}, \citenamefont {Parschau},
  \citenamefont {Macia}, \citenamefont {Katsonis}, \citenamefont {Harutyunyan},
  \citenamefont {Ernst},\ and\ \citenamefont {Feringa}}]{nanocar}%
  \BibitemOpen
  \bibfield  {author} {\bibinfo {author} {\bibfnamefont {T.}~\bibnamefont
  {Kudernac}}, \bibinfo {author} {\bibfnamefont {N.}~\bibnamefont
  {Ruangsupapichat}}, \bibinfo {author} {\bibfnamefont {M.}~\bibnamefont
  {Parschau}}, \bibinfo {author} {\bibfnamefont {B.}~\bibnamefont {Macia}},
  \bibinfo {author} {\bibfnamefont {N.}~\bibnamefont {Katsonis}}, \bibinfo
  {author} {\bibfnamefont {S.~R.}\ \bibnamefont {Harutyunyan}}, \bibinfo
  {author} {\bibfnamefont {K.-H.}\ \bibnamefont {Ernst}}, \ and\ \bibinfo
  {author} {\bibfnamefont {B.~L.}\ \bibnamefont {Feringa}},\ }\href@noop {}
  {\bibfield  {journal} {\bibinfo  {journal} {Nature}\ }\textbf {\bibinfo
  {volume} {479}},\ \bibinfo {pages} {208} (\bibinfo {year}
  {2011})}\BibitemShut {NoStop}%
\bibitem [{\citenamefont {Tierney}\ \emph {et~al.}(2011)\citenamefont
  {Tierney}, \citenamefont {Murphy}, \citenamefont {Jewell}, \citenamefont
  {Baber}, \citenamefont {Iski}, \citenamefont {Khodaverdian}, \citenamefont
  {McGuire}, \citenamefont {Klebanov},\ and\ \citenamefont
  {Sykes}}]{NatureNano11}%
  \BibitemOpen
  \bibfield  {author} {\bibinfo {author} {\bibfnamefont {H.~L.}\ \bibnamefont
  {Tierney}}, \bibinfo {author} {\bibfnamefont {C.~J.}\ \bibnamefont {Murphy}},
  \bibinfo {author} {\bibfnamefont {A.~D.}\ \bibnamefont {Jewell}}, \bibinfo
  {author} {\bibfnamefont {A.~E.}\ \bibnamefont {Baber}}, \bibinfo {author}
  {\bibfnamefont {E.~V.}\ \bibnamefont {Iski}}, \bibinfo {author}
  {\bibfnamefont {H.~Y.}\ \bibnamefont {Khodaverdian}}, \bibinfo {author}
  {\bibfnamefont {A.~F.}\ \bibnamefont {McGuire}}, \bibinfo {author}
  {\bibfnamefont {N.}~\bibnamefont {Klebanov}}, \ and\ \bibinfo {author}
  {\bibfnamefont {E.~C.~H.}\ \bibnamefont {Sykes}},\ }\href@noop {} {\bibfield
  {journal} {\bibinfo  {journal} {Nature nanotechnology}\ }\textbf {\bibinfo
  {volume} {6}},\ \bibinfo {pages} {625} (\bibinfo {year} {2011})}\BibitemShut
  {NoStop}%
\bibitem [{\citenamefont {Chiaravalloti}\ \emph {et~al.}(2007)\citenamefont
  {Chiaravalloti}, \citenamefont {Gross}, \citenamefont {Rieder}, \citenamefont
  {Stojkovic}, \citenamefont {Gourdon}, \citenamefont {Joachim},\ and\
  \citenamefont {Moresco}}]{chiaravalloti2007rack}%
  \BibitemOpen
  \bibfield  {author} {\bibinfo {author} {\bibfnamefont {F.}~\bibnamefont
  {Chiaravalloti}}, \bibinfo {author} {\bibfnamefont {L.}~\bibnamefont
  {Gross}}, \bibinfo {author} {\bibfnamefont {K.-H.}\ \bibnamefont {Rieder}},
  \bibinfo {author} {\bibfnamefont {S.~M.}\ \bibnamefont {Stojkovic}}, \bibinfo
  {author} {\bibfnamefont {A.}~\bibnamefont {Gourdon}}, \bibinfo {author}
  {\bibfnamefont {C.}~\bibnamefont {Joachim}}, \ and\ \bibinfo {author}
  {\bibfnamefont {F.}~\bibnamefont {Moresco}},\ }\href@noop {} {\bibfield
  {journal} {\bibinfo  {journal} {Nature Materials}\ }\textbf {\bibinfo
  {volume} {6}},\ \bibinfo {pages} {30} (\bibinfo {year} {2007})}\BibitemShut
  {NoStop}%
\bibitem [{\citenamefont {Bustos-Mar\'un}\ \emph {et~al.}(2013)\citenamefont
  {Bustos-Mar\'un}, \citenamefont {Refael},\ and\ \citenamefont {von
  Oppen}}]{AQM13}%
  \BibitemOpen
  \bibfield  {author} {\bibinfo {author} {\bibfnamefont {R.}~\bibnamefont
  {Bustos-Mar\'un}}, \bibinfo {author} {\bibfnamefont {G.}~\bibnamefont
  {Refael}}, \ and\ \bibinfo {author} {\bibfnamefont {F.}~\bibnamefont {von
  Oppen}},\ }\href@noop {} {\bibfield  {journal} {\bibinfo  {journal} {Phys.
  Rev. Lett.}\ }\textbf {\bibinfo {volume} {111}},\ \bibinfo {pages} {060802}
  (\bibinfo {year} {2013})}\BibitemShut {NoStop}%
\bibitem [{\citenamefont {Fern\'andez-Alc\'azar}\ \emph
  {et~al.}(2015)\citenamefont {Fern\'andez-Alc\'azar}, \citenamefont
  {Bustos-Mar\'un},\ and\ \citenamefont {Pastawski}}]{FBP15}%
  \BibitemOpen
  \bibfield  {author} {\bibinfo {author} {\bibfnamefont {L.~J.}\ \bibnamefont
  {Fern\'andez-Alc\'azar}}, \bibinfo {author} {\bibfnamefont {R.~A.}\
  \bibnamefont {Bustos-Mar\'un}}, \ and\ \bibinfo {author} {\bibfnamefont
  {H.~M.}\ \bibnamefont {Pastawski}},\ }\href {\doibase
  10.1103/PhysRevB.92.075406} {\bibfield  {journal} {\bibinfo  {journal} {Phys.
  Rev. B}\ }\textbf {\bibinfo {volume} {92}},\ \bibinfo {pages} {075406}
  (\bibinfo {year} {2015})}\BibitemShut {NoStop}%
\bibitem [{\citenamefont {Arrachea}\ and\ \citenamefont {von
  Oppen}(2015)}]{arrachea2015}%
  \BibitemOpen
  \bibfield  {author} {\bibinfo {author} {\bibfnamefont {L.}~\bibnamefont
  {Arrachea}}\ and\ \bibinfo {author} {\bibfnamefont {F.}~\bibnamefont {von
  Oppen}},\ }\href {\doibase 10.1016/j.physe.2015.08.031} {\bibfield  {journal}
  {\bibinfo  {journal} {Physica E}\ }\textbf {\bibinfo {volume} {74}},\
  \bibinfo {pages} {96} (\bibinfo {year} {2015})}\BibitemShut {NoStop}%
\bibitem [{\citenamefont {Silvestrov}\ \emph {et~al.}(2016)\citenamefont
  {Silvestrov}, \citenamefont {Recher},\ and\ \citenamefont
  {Brouwer}}]{Silvestrov2016}%
  \BibitemOpen
  \bibfield  {author} {\bibinfo {author} {\bibfnamefont {P.~G.}\ \bibnamefont
  {Silvestrov}}, \bibinfo {author} {\bibfnamefont {P.}~\bibnamefont {Recher}},
  \ and\ \bibinfo {author} {\bibfnamefont {P.~W.}\ \bibnamefont {Brouwer}},\
  }\href {\doibase 10.1103/PhysRevB.93.205130} {\bibfield  {journal} {\bibinfo
  {journal} {Phys. Rev. B}\ }\textbf {\bibinfo {volume} {93}},\ \bibinfo
  {pages} {205130} (\bibinfo {year} {2016})}\BibitemShut {NoStop}%
\bibitem [{\citenamefont {Ludovico}\ \emph {et~al.}(2016)\citenamefont
  {Ludovico}, \citenamefont {Battista}, \citenamefont {von Oppen},\ and\
  \citenamefont {Arrachea}}]{Liliana_Onsager}%
  \BibitemOpen
  \bibfield  {author} {\bibinfo {author} {\bibfnamefont {M.~F.}\ \bibnamefont
  {Ludovico}}, \bibinfo {author} {\bibfnamefont {F.}~\bibnamefont {Battista}},
  \bibinfo {author} {\bibfnamefont {F.}~\bibnamefont {von Oppen}}, \ and\
  \bibinfo {author} {\bibfnamefont {L.}~\bibnamefont {Arrachea}},\ }\href
  {\doibase 10.1103/PhysRevB.93.075136} {\bibfield  {journal} {\bibinfo
  {journal} {Phys. Rev. B}\ }\textbf {\bibinfo {volume} {93}},\ \bibinfo
  {pages} {075136} (\bibinfo {year} {2016})}\BibitemShut {NoStop}%
\bibitem [{\citenamefont {Calvo}\ \emph {et~al.}(2017)\citenamefont {Calvo},
  \citenamefont {Ribetto},\ and\ \citenamefont {Bustos-Mar\'un}}]{calvo2017}%
  \BibitemOpen
  \bibfield  {author} {\bibinfo {author} {\bibfnamefont {H.~L.}\ \bibnamefont
  {Calvo}}, \bibinfo {author} {\bibfnamefont {F.~D.}\ \bibnamefont {Ribetto}},
  \ and\ \bibinfo {author} {\bibfnamefont {R.~A.}\ \bibnamefont
  {Bustos-Mar\'un}},\ }\href@noop {} {\bibfield  {journal} {\bibinfo  {journal}
  {Phys. Rev. B}\ }\textbf {\bibinfo {volume} {96}},\ \bibinfo {pages} {165309}
  (\bibinfo {year} {2017})}\BibitemShut {NoStop}%
\bibitem [{\citenamefont {Landauer}\ and\ \citenamefont
  {Woo}(1974)}]{landauer74driving}%
  \BibitemOpen
  \bibfield  {author} {\bibinfo {author} {\bibfnamefont {R.}~\bibnamefont
  {Landauer}}\ and\ \bibinfo {author} {\bibfnamefont {J.~W.~F.}\ \bibnamefont
  {Woo}},\ }\href@noop {} {\bibfield  {journal} {\bibinfo  {journal} {Physical
  Review B}\ }\textbf {\bibinfo {volume} {10}},\ \bibinfo {pages} {1266}
  (\bibinfo {year} {1974})}\BibitemShut {NoStop}%
\bibitem [{\citenamefont {Thouless}(1983)}]{Thouless83}%
  \BibitemOpen
  \bibfield  {author} {\bibinfo {author} {\bibfnamefont {D.~J.}\ \bibnamefont
  {Thouless}},\ }\href@noop {} {\bibfield  {journal} {\bibinfo  {journal}
  {Physical Review B}\ }\textbf {\bibinfo {volume} {27}},\ \bibinfo {pages}
  {6083} (\bibinfo {year} {1983})}\BibitemShut {NoStop}%
\bibitem [{\citenamefont {Qi}\ and\ \citenamefont {Zhang}(2009)}]{Zhang}%
  \BibitemOpen
  \bibfield  {author} {\bibinfo {author} {\bibfnamefont {X.-L.}\ \bibnamefont
  {Qi}}\ and\ \bibinfo {author} {\bibfnamefont {S.~C.}\ \bibnamefont {Zhang}},\
  }\href@noop {} {\bibfield  {journal} {\bibinfo  {journal} {Phys. Rev. B}\
  }\textbf {\bibinfo {volume} {79}},\ \bibinfo {pages} {235442} (\bibinfo
  {year} {2009})}\BibitemShut {NoStop}%
\bibitem [{Note1()}]{Note1}%
  \BibitemOpen
  \bibinfo {note} {It has been shown that localization can improve charge
  pumping. See for example Refs. \cite {chern2007} and \cite
  {ingaramo2013}}\BibitemShut {NoStop}%
\bibitem [{Note2()}]{Note2}%
  \BibitemOpen
  \bibinfo {note} {While this work was under the reviewing process, we found a
  work with a related idea, but about the possibility of using
  many-body-localization to make an engine. N. Y. Halpern, C. D. White, S.
  Gopalakrishnan, and G. Refael . MBL-mobile: Many-body-localized engine.
  arXiv:1707.07008 (2018)}\BibitemShut {NoStop}%
\bibitem [{\citenamefont {Das}\ and\ \citenamefont
  {Peierls}(1975)}]{Peierls1975force}%
  \BibitemOpen
  \bibfield  {author} {\bibinfo {author} {\bibfnamefont {A.~K.}\ \bibnamefont
  {Das}}\ and\ \bibinfo {author} {\bibfnamefont {R.~E.}\ \bibnamefont
  {Peierls}},\ }\href@noop {} {\bibfield  {journal} {\bibinfo  {journal}
  {Journal of Physics C: Solid State Physics}\ }\textbf {\bibinfo {volume}
  {8}},\ \bibinfo {pages} {3348} (\bibinfo {year} {1975})}\BibitemShut
  {NoStop}%
\bibitem [{\citenamefont {G{\'o}mez-Medina}\ \emph {et~al.}(2001)\citenamefont
  {G{\'o}mez-Medina}, \citenamefont {San~Jos{\'e}}, \citenamefont
  {Garc{\'\i}a-Mart{\'\i}n}, \citenamefont {Lester}, \citenamefont
  {Nieto-Vesperinas},\ and\ \citenamefont {S{\'a}enz}}]{Saenz01}%
  \BibitemOpen
  \bibfield  {author} {\bibinfo {author} {\bibfnamefont {R.}~\bibnamefont
  {G{\'o}mez-Medina}}, \bibinfo {author} {\bibfnamefont {P.}~\bibnamefont
  {San~Jos{\'e}}}, \bibinfo {author} {\bibfnamefont {A.}~\bibnamefont
  {Garc{\'\i}a-Mart{\'\i}n}}, \bibinfo {author} {\bibfnamefont
  {M.}~\bibnamefont {Lester}}, \bibinfo {author} {\bibfnamefont
  {M.}~\bibnamefont {Nieto-Vesperinas}}, \ and\ \bibinfo {author}
  {\bibfnamefont {J.~J.}\ \bibnamefont {S{\'a}enz}},\ }\href@noop {} {\bibfield
   {journal} {\bibinfo  {journal} {Physical review letters}\ }\textbf {\bibinfo
  {volume} {86}},\ \bibinfo {pages} {4275} (\bibinfo {year}
  {2001})}\BibitemShut {NoStop}%
\bibitem [{\citenamefont {Bennett}\ \emph {et~al.}(2010)\citenamefont
  {Bennett}, \citenamefont {Maassen},\ and\ \citenamefont
  {Clerk}}]{bennett2010}%
  \BibitemOpen
  \bibfield  {author} {\bibinfo {author} {\bibfnamefont {S.~D.}\ \bibnamefont
  {Bennett}}, \bibinfo {author} {\bibfnamefont {J.}~\bibnamefont {Maassen}}, \
  and\ \bibinfo {author} {\bibfnamefont {A.~A.}\ \bibnamefont {Clerk}},\ }\href
  {\doibase 10.1103/PhysRevLett.105.217206} {\bibfield  {journal} {\bibinfo
  {journal} {Phys. Rev. Lett.}\ }\textbf {\bibinfo {volume} {105}},\ \bibinfo
  {pages} {217206} (\bibinfo {year} {2010})}\BibitemShut {NoStop}%
\bibitem [{\citenamefont {Thomas}\ \emph {et~al.}(2012)\citenamefont {Thomas},
  \citenamefont {Karzig}, \citenamefont {Viola~Kusminskiy}, \citenamefont
  {Zar\'and},\ and\ \citenamefont {von Oppen}}]{thomas2012}%
  \BibitemOpen
  \bibfield  {author} {\bibinfo {author} {\bibfnamefont {M.}~\bibnamefont
  {Thomas}}, \bibinfo {author} {\bibfnamefont {T.}~\bibnamefont {Karzig}},
  \bibinfo {author} {\bibfnamefont {S.}~\bibnamefont {Viola~Kusminskiy}},
  \bibinfo {author} {\bibfnamefont {G.}~\bibnamefont {Zar\'and}}, \ and\
  \bibinfo {author} {\bibfnamefont {F.}~\bibnamefont {von Oppen}},\ }\href
  {\doibase 10.1103/PhysRevB.86.195419} {\bibfield  {journal} {\bibinfo
  {journal} {Phys. Rev. B}\ }\textbf {\bibinfo {volume} {86}},\ \bibinfo
  {pages} {195419} (\bibinfo {year} {2012})}\BibitemShut {NoStop}%
\bibitem [{\citenamefont {Di~Ventra}\ and\ \citenamefont
  {Pantelides}(2000)}]{diventra2000}%
  \BibitemOpen
  \bibfield  {author} {\bibinfo {author} {\bibfnamefont {M.}~\bibnamefont
  {Di~Ventra}}\ and\ \bibinfo {author} {\bibfnamefont {S.~T.}\ \bibnamefont
  {Pantelides}},\ }\href {\doibase 10.1103/PhysRevB.61.16207} {\bibfield
  {journal} {\bibinfo  {journal} {Phys. Rev. B}\ }\textbf {\bibinfo {volume}
  {61}},\ \bibinfo {pages} {16207} (\bibinfo {year} {2000})}\BibitemShut
  {NoStop}%
\bibitem [{\citenamefont {Horsfield}\ \emph {et~al.}(2004)\citenamefont
  {Horsfield}, \citenamefont {Bowler}, \citenamefont {Fisher}, \citenamefont
  {Todorov},\ and\ \citenamefont {S\'anchez}}]{horsfield2004}%
  \BibitemOpen
  \bibfield  {author} {\bibinfo {author} {\bibfnamefont {A.~P.}\ \bibnamefont
  {Horsfield}}, \bibinfo {author} {\bibfnamefont {D.~R.}\ \bibnamefont
  {Bowler}}, \bibinfo {author} {\bibfnamefont {A.~J.}\ \bibnamefont {Fisher}},
  \bibinfo {author} {\bibfnamefont {T.~N.}\ \bibnamefont {Todorov}}, \ and\
  \bibinfo {author} {\bibfnamefont {C.~G.}\ \bibnamefont {S\'anchez}},\ }\href
  {\doibase 10.1088/0953-8984/16/46/012} {\bibfield  {journal} {\bibinfo
  {journal} {J. Phys. Condens. Matter}\ }\textbf {\bibinfo {volume} {16}},\
  \bibinfo {pages} {8251} (\bibinfo {year} {2004})}\BibitemShut {NoStop}%
\bibitem [{\citenamefont {Todorov}\ and\ \citenamefont
  {Dundas}(2010)}]{todorov2010}%
  \BibitemOpen
  \bibfield  {author} {\bibinfo {author} {\bibfnamefont {T.~N.}\ \bibnamefont
  {Todorov}}\ and\ \bibinfo {author} {\bibfnamefont {D.}~\bibnamefont
  {Dundas}},\ }\href {\doibase 10.1103/PhysRevB.81.075416} {\bibfield
  {journal} {\bibinfo  {journal} {Phys. Rev. B}\ }\textbf {\bibinfo {volume}
  {81}},\ \bibinfo {pages} {075416} (\bibinfo {year} {2010})}\BibitemShut
  {NoStop}%
\bibitem [{\citenamefont {Haug}\ and\ \citenamefont {Jauho}(2008)}]{haug2008}%
  \BibitemOpen
  \bibfield  {author} {\bibinfo {author} {\bibfnamefont {H.}~\bibnamefont
  {Haug}}\ and\ \bibinfo {author} {\bibfnamefont {A.-P.}\ \bibnamefont
  {Jauho}},\ }\href {\doibase 10.1007/978-3-540-73564-9} {\emph {\bibinfo
  {title} {Quantum kinetics in transport and optics of semiconductors}}},\
  \bibinfo {edition} {2nd}\ ed.,\ Solid-State Sciences 123\ (\bibinfo
  {publisher} {Springer-Verlag Berlin Heidelberg},\ \bibinfo {year}
  {2008})\BibitemShut {NoStop}%
\bibitem [{\citenamefont {{Pastawski}}(1992)}]{GLBE2}%
  \BibitemOpen
  \bibfield  {author} {\bibinfo {author} {\bibfnamefont {H.~M.}\ \bibnamefont
  {{Pastawski}}},\ }\href@noop {} {\bibfield  {journal} {\bibinfo  {journal}
  {Phys. Rev. B}\ }\textbf {\bibinfo {volume} {46}},\ \bibinfo {pages} {4053}
  (\bibinfo {year} {1992})}\BibitemShut {NoStop}%
\bibitem [{\citenamefont {Bode}\ \emph {et~al.}(2011)\citenamefont {Bode},
  \citenamefont {Viola~Kusminskiy}, \citenamefont {Egger},\ and\ \citenamefont
  {von Oppen}}]{bode2011}%
  \BibitemOpen
  \bibfield  {author} {\bibinfo {author} {\bibfnamefont {N.}~\bibnamefont
  {Bode}}, \bibinfo {author} {\bibfnamefont {S.}~\bibnamefont
  {Viola~Kusminskiy}}, \bibinfo {author} {\bibfnamefont {R.}~\bibnamefont
  {Egger}}, \ and\ \bibinfo {author} {\bibfnamefont {F.}~\bibnamefont {von
  Oppen}},\ }\href {\doibase 10.1103/PhysRevLett.107.036804} {\bibfield
  {journal} {\bibinfo  {journal} {Phys. Rev. Lett.}\ }\textbf {\bibinfo
  {volume} {107}},\ \bibinfo {pages} {036804} (\bibinfo {year}
  {2011})}\BibitemShut {NoStop}%
\bibitem [{\citenamefont {Bode}\ \emph {et~al.}(2012)\citenamefont {Bode},
  \citenamefont {Viola~Kusminskiy}, \citenamefont {Egger},\ and\ \citenamefont
  {von Oppen}}]{bode2012}%
  \BibitemOpen
  \bibfield  {author} {\bibinfo {author} {\bibfnamefont {N.}~\bibnamefont
  {Bode}}, \bibinfo {author} {\bibfnamefont {S.}~\bibnamefont
  {Viola~Kusminskiy}}, \bibinfo {author} {\bibfnamefont {R.}~\bibnamefont
  {Egger}}, \ and\ \bibinfo {author} {\bibfnamefont {F.}~\bibnamefont {von
  Oppen}},\ }\href@noop {} {\bibfield  {journal} {\bibinfo  {journal}
  {Beilstein J. Nanotechnol.}\ }\textbf {\bibinfo {volume} {3}},\ \bibinfo
  {pages} {144} (\bibinfo {year} {2012})}\BibitemShut {NoStop}%
\bibitem [{\citenamefont {Brouwer}(1998)}]{brouwer1998}%
  \BibitemOpen
  \bibfield  {author} {\bibinfo {author} {\bibfnamefont {P.~W.}\ \bibnamefont
  {Brouwer}},\ }\href@noop {} {\bibfield  {journal} {\bibinfo  {journal} {Phys.
  Rev. B}\ }\textbf {\bibinfo {volume} {58}},\ \bibinfo {pages} {R10135}
  (\bibinfo {year} {1998})}\BibitemShut {NoStop}%
\bibitem [{\citenamefont {Avron}\ \emph {et~al.}(2004)\citenamefont {Avron},
  \citenamefont {Elgart}, \citenamefont {Graf},\ and\ \citenamefont
  {Sadun}}]{Avron2004}%
  \BibitemOpen
  \bibfield  {author} {\bibinfo {author} {\bibfnamefont {J.~E.}\ \bibnamefont
  {Avron}}, \bibinfo {author} {\bibfnamefont {A.}~\bibnamefont {Elgart}},
  \bibinfo {author} {\bibfnamefont {G.~M.}\ \bibnamefont {Graf}}, \ and\
  \bibinfo {author} {\bibfnamefont {L.}~\bibnamefont {Sadun}},\ }\href@noop {}
  {\bibfield  {journal} {\bibinfo  {journal} {Journal of Statistical Physics}\
  }\textbf {\bibinfo {volume} {116}},\ \bibinfo {pages} {425} (\bibinfo {year}
  {2004})}\BibitemShut {NoStop}%
\bibitem [{Note3()}]{Note3}%
  \BibitemOpen
  \bibinfo {note} {As in the case of the Thouless motor\cite {FPB17}, in the
  rotor, weak potentials and the smoothening of the system's edges make
  $F^{eq}$ small as compared with $F^{neq}$. For the shuttle, $F^{eq}$ comes
  from imperfections of the device, which we are neglecting in the present
  work.}\BibitemShut {Stop}%
\bibitem [{\citenamefont {Cohen}(2003)}]{Cohen}%
  \BibitemOpen
  \bibfield  {author} {\bibinfo {author} {\bibfnamefont {D.}~\bibnamefont
  {Cohen}},\ }\href@noop {} {\bibfield  {journal} {\bibinfo  {journal} {Phys.
  Rev. B}\ }\textbf {\bibinfo {volume} {68}},\ \bibinfo {pages} {201303R}
  (\bibinfo {year} {2003})}\BibitemShut {NoStop}%
\bibitem [{\citenamefont {Cohen}\ \emph {et~al.}(2005)\citenamefont {Cohen},
  \citenamefont {Kottos},\ and\ \citenamefont {Schanz}}]{Cohen2005}%
  \BibitemOpen
  \bibfield  {author} {\bibinfo {author} {\bibfnamefont {D.}~\bibnamefont
  {Cohen}}, \bibinfo {author} {\bibfnamefont {T.}~\bibnamefont {Kottos}}, \
  and\ \bibinfo {author} {\bibfnamefont {H.}~\bibnamefont {Schanz}},\
  }\href@noop {} {\bibfield  {journal} {\bibinfo  {journal} {Phys. Rev. E}\
  }\textbf {\bibinfo {volume} {71}},\ \bibinfo {pages} {035202} (\bibinfo
  {year} {2005})}\BibitemShut {NoStop}%
\bibitem [{\citenamefont {Anderson}(1978)}]{Anderson78}%
  \BibitemOpen
  \bibfield  {author} {\bibinfo {author} {\bibfnamefont {P.~W.}\ \bibnamefont
  {Anderson}},\ }\href@noop {} {\bibfield  {journal} {\bibinfo  {journal} {Rev.
  Mod. Phys.}\ }\textbf {\bibinfo {volume} {50}},\ \bibinfo {pages} {191}
  (\bibinfo {year} {1978})}\BibitemShut {NoStop}%
\bibitem [{\citenamefont {Pastawski}\ \emph {et~al.}(1983)\citenamefont
  {Pastawski}, \citenamefont {Weisz},\ and\ \citenamefont {Albornoz}}]{PWA83}%
  \BibitemOpen
  \bibfield  {author} {\bibinfo {author} {\bibfnamefont {H.~M.}\ \bibnamefont
  {Pastawski}}, \bibinfo {author} {\bibfnamefont {J.~F.}\ \bibnamefont
  {Weisz}}, \ and\ \bibinfo {author} {\bibfnamefont {S.}~\bibnamefont
  {Albornoz}},\ }\href {\doibase 10.1103/PhysRevB.28.6896} {\bibfield
  {journal} {\bibinfo  {journal} {Phys. Rev. B}\ }\textbf {\bibinfo {volume}
  {28}},\ \bibinfo {pages} {6896} (\bibinfo {year} {1983})}\BibitemShut
  {NoStop}%
\bibitem [{\citenamefont {Pastawski}\ \emph {et~al.}(1985)\citenamefont
  {Pastawski}, \citenamefont {Slutzky},\ and\ \citenamefont
  {Weisz}}]{PSW85_breakdown}%
  \BibitemOpen
  \bibfield  {author} {\bibinfo {author} {\bibfnamefont {H.~M.}\ \bibnamefont
  {Pastawski}}, \bibinfo {author} {\bibfnamefont {C.~M.}\ \bibnamefont
  {Slutzky}}, \ and\ \bibinfo {author} {\bibfnamefont {J.~F.}\ \bibnamefont
  {Weisz}},\ }\href {\doibase 10.1103/PhysRevB.32.3642} {\bibfield  {journal}
  {\bibinfo  {journal} {Phys. Rev. B}\ }\textbf {\bibinfo {volume} {32}},\
  \bibinfo {pages} {3642} (\bibinfo {year} {1985})}\BibitemShut {NoStop}%
\bibitem [{\citenamefont {Kramer}\ and\ \citenamefont
  {MacKinnon}(1993)}]{KMK1993}%
  \BibitemOpen
  \bibfield  {author} {\bibinfo {author} {\bibfnamefont {B.}~\bibnamefont
  {Kramer}}\ and\ \bibinfo {author} {\bibfnamefont {A.}~\bibnamefont
  {MacKinnon}},\ }\href@noop {} {\bibfield  {journal} {\bibinfo  {journal}
  {Reports on Progress in Physics}\ }\textbf {\bibinfo {volume} {56}},\
  \bibinfo {pages} {1469} (\bibinfo {year} {1993})}\BibitemShut {NoStop}%
\bibitem [{\citenamefont {Pastawski}\ and\ \citenamefont
  {Medina}(2001)}]{PM01}%
  \BibitemOpen
  \bibfield  {author} {\bibinfo {author} {\bibfnamefont {H.~M.}\ \bibnamefont
  {Pastawski}}\ and\ \bibinfo {author} {\bibfnamefont {E.}~\bibnamefont
  {Medina}},\ }\href@noop {} {\bibfield  {journal} {\bibinfo  {journal} {Rev.
  Mex. Fis.}\ }\textbf {\bibinfo {volume} {47}},\ \bibinfo {pages} {1}
  (\bibinfo {year} {2001})}\BibitemShut {NoStop}%
\bibitem [{Note4()}]{Note4}%
  \BibitemOpen
  \bibinfo {note} {In all the figures of this manuscript, we used Eq. \ref
  {eq:P_gamma} to calculate the probability distribution function of the
  properties of interest. For other types of disorder, alternative expressions
  for Eq. \ref {eq:P_gamma} may hold. However, the expressions for the
  probability distribution functions of the force (Eq. \ref {eq:P_F}) and the
  efficiency (Eq. \ref {eq:P_eta_AQM}) are written in such a way that they are
  independent of the particular form of the probability distribution function
  of $\xi $.}\BibitemShut {Stop}%
\bibitem [{Note5()}]{Note5}%
  \BibitemOpen
  \bibinfo {note} {The ensemble described by the probability distribution
  functions may consist of different realizations of the system but may also
  come from evaluating the same system at different Fermi energies, which
  should involve distinct sets of localized states.}\BibitemShut {Stop}%
\bibitem [{\citenamefont {Van~Kampen}(1992)}]{vanKampen}%
  \BibitemOpen
  \bibfield  {author} {\bibinfo {author} {\bibfnamefont {N.~G.}\ \bibnamefont
  {Van~Kampen}},\ }\href@noop {} {\emph {\bibinfo {title} {Stochastic processes
  in physics and chemistry}}}\ (\bibinfo  {publisher} {Elsevier, Amsterdam, The
  Netherlands},\ \bibinfo {year} {1992})\BibitemShut {NoStop}%
\bibitem [{\citenamefont {Weisz}\ and\ \citenamefont {Pastawski}(1984)}]{WP82}%
  \BibitemOpen
  \bibfield  {author} {\bibinfo {author} {\bibfnamefont {J.~F.}\ \bibnamefont
  {Weisz}}\ and\ \bibinfo {author} {\bibfnamefont {H.~M.}\ \bibnamefont
  {Pastawski}},\ }\href@noop {} {\bibfield  {journal} {\bibinfo  {journal}
  {Phys. Lett. A}\ }\textbf {\bibinfo {volume} {105}},\ \bibinfo {pages} {421}
  (\bibinfo {year} {1984})}\BibitemShut {NoStop}%
\bibitem [{Note6()}]{Note6}%
  \BibitemOpen
  \bibinfo {note} {The values of $v_F$ and $k_F$ were taken from the order of
  magnitude of typical metals.\cite {AshcroftMermin} The value of $\gamma $
  used ($2.5 \times 10^{-8} kg/s$) implies $\tau = 1 \times 10^{-2}s$ at
  $\delta \mu = 1 \times 10^{-3} eV$. This was estimated from $\delta \mu
  k_F/\pi = \gamma \protect \mathaccentV {dot}05Fx$ and assuming a constant
  terminal velocity with $F^{load} = 0$. The value of $\tau = 1 \times
  10^{-2}s$ is of the order of the nanomotor reported by K. Kim et. al.\cite
  {FastNM}.}\BibitemShut {Stop}%
\bibitem [{\citenamefont {Cattena}\ \emph {et~al.}(2010)\citenamefont
  {Cattena}, \citenamefont {Bustos-Mar\'{u}n},\ and\ \citenamefont
  {Pastawski}}]{CBMP10}%
  \BibitemOpen
  \bibfield  {author} {\bibinfo {author} {\bibfnamefont {C.~J.}\ \bibnamefont
  {Cattena}}, \bibinfo {author} {\bibfnamefont {R.~A.}\ \bibnamefont
  {Bustos-Mar\'{u}n}}, \ and\ \bibinfo {author} {\bibfnamefont {H.~M.}\
  \bibnamefont {Pastawski}},\ }\href@noop {} {\bibfield  {journal} {\bibinfo
  {journal} {Phys. Rev. B}\ }\textbf {\bibinfo {volume} {82}},\ \bibinfo
  {pages} {144201} (\bibinfo {year} {2010})}\BibitemShut {NoStop}%
\bibitem [{\citenamefont {Ashcroft}\ and\ \citenamefont
  {Mermin}(1976)}]{AshcroftMermin}%
  \BibitemOpen
  \bibfield  {author} {\bibinfo {author} {\bibfnamefont {N.~W.}\ \bibnamefont
  {Ashcroft}}\ and\ \bibinfo {author} {\bibfnamefont {N.~D.}\ \bibnamefont
  {Mermin}},\ }\href@noop {} {\emph {\bibinfo {title} {Solid state physics}}}\
  (\bibinfo  {publisher} {Saunders College, Philadelphia},\ \bibinfo {year}
  {1976})\BibitemShut {NoStop}%
\bibitem [{\citenamefont {Nazarov}\ and\ \citenamefont
  {Blanter}(2009)}]{Nazarov2009}%
  \BibitemOpen
  \bibfield  {author} {\bibinfo {author} {\bibfnamefont {Y.}~\bibnamefont
  {Nazarov}}\ and\ \bibinfo {author} {\bibfnamefont {Y.}~\bibnamefont
  {Blanter}},\ }\href@noop {} {\emph {\bibinfo {title} {Quantum Transport.
  Introduction to Nanoscience}}}\ (\bibinfo  {publisher} {Cambridge University
  Press, New York},\ \bibinfo {year} {2009})\BibitemShut {NoStop}%
\bibitem [{\citenamefont {Chern}\ \emph {et~al.}(2007)\citenamefont {Chern},
  \citenamefont {Onoda}, \citenamefont {Murakami},\ and\ \citenamefont
  {Nagaosa}}]{chern2007}%
  \BibitemOpen
  \bibfield  {author} {\bibinfo {author} {\bibfnamefont {C.-H.}\ \bibnamefont
  {Chern}}, \bibinfo {author} {\bibfnamefont {S.}~\bibnamefont {Onoda}},
  \bibinfo {author} {\bibfnamefont {S.}~\bibnamefont {Murakami}}, \ and\
  \bibinfo {author} {\bibfnamefont {N.}~\bibnamefont {Nagaosa}},\ }\href
  {\doibase 10.1103/PhysRevB.76.035334} {\bibfield  {journal} {\bibinfo
  {journal} {Phys. Rev. B}\ }\textbf {\bibinfo {volume} {76}},\ \bibinfo
  {pages} {035334} (\bibinfo {year} {2007})}\BibitemShut {NoStop}%
\bibitem [{\citenamefont {Ingaramo}\ and\ \citenamefont
  {Foa~Torres}(2013)}]{ingaramo2013}%
  \BibitemOpen
  \bibfield  {author} {\bibinfo {author} {\bibfnamefont {L.~H.}\ \bibnamefont
  {Ingaramo}}\ and\ \bibinfo {author} {\bibfnamefont {L.~E.~F.}\ \bibnamefont
  {Foa~Torres}},\ }\href {\doibase 10.1063/1.4821262} {\bibfield  {journal}
  {\bibinfo  {journal} {Appl. Phys. Lett.}\ }\textbf {\bibinfo {volume}
  {103}},\ \bibinfo {pages} {123508} (\bibinfo {year} {2013})}\BibitemShut
  {NoStop}%
\end{thebibliography}%

\end{document}